\documentclass[a4paper,11pt]{article}
\usepackage{jheppub} 
\usepackage{lineno}
\usepackage[T1]{fontenc} 
\usepackage{slashed}

\usepackage{multicol}
\usepackage{braket}
\usepackage[dvipsnames]{xcolor}
\usepackage{comment}
\usepackage{float}
\usepackage{caption}
\usepackage{subcaption}
\usepackage{bbold}  
\usepackage{hyperref}
\usepackage{multirow}
\title{\boldmath On deformations of AdS$_3$ solutions, supersymmetry and G-structures}

\author[a]{Anayeli Ram\'irez}
\author[b]{and Salom\'on Zacar\'ias}

\affiliation[a]{Instituut voor Theoretische Fysica, KU Leuven,
\\
Celestijnenlaan 200D, 3000 Leuven, Belgium
}
\affiliation[b]{Department of Theoretical Physics and Astrophysics, Faculty of Science, Masaryk University,\\
611 37 Brno, Czech Republic}

\emailAdd{ mariaanayeli.ramirezortiz@kuleuven.be, szacarias@physics.muni.cz,}

\abstract{We construct new families of supersymmetric AdS$_3$ solutions in both massive and massless Type IIA supergravity via deformations to known backgrounds preserving $\mathcal{N} = (4,0)$ and $\mathcal{N} = (6,0)$ supersymmetry. These deformations are performed along internal isometries and lead to backgrounds with fully preserved or reduced supersymmetry. Using the formalism of $G$-structures, we systematically characterise the resulting geometries and track the evolution of their supersymmetric properties under the deformation. In particular, we identify transitions among SU(3), SU(2), and identity structures, and demonstrate the preservation of Killing spinors through explicit spinor bilinear constructions. Additionally, we investigate D-brane embeddings in the deformed geometries, uncovering stable and supersymmetric configurations supported by the new backgrounds. Our results offer new insights into the classification of AdS$_3$ flux vacua and provide a concrete framework for understanding their potential holographic duals.
}

\begin{document}
\maketitle
\flushbottom
\providecommand{\abs}[1]{\lvert#1\rvert}

\section{Introduction}
\label{sec:intro}


The study of the AdS$_3$/CFT$_2$ correspondence has long stood as a cornerstone of string theory, offering one of the most tractable and well-understood realisations of the holographic principle. Its tractability stems from the  propierties of two-dimensional conformal field theories (CFTs), which are particularly amenable to exact analyses owing to the infinite-dimensional nature of their conformal symmetry algebra. This rich symmetry structure renders AdS$_3$/CFT$_2$ dualities more analytically accessible than their higher-dimensional counterparts.

A prominent and extensively studied example of AdS$_3$ holography arises in the near-horizon limit of D1-D5 brane configurations, which give rise to geometries of the form AdS$_3 \times$S$^3 \times$CY$_2$, where CY$_2$ is either T$^4$ or K3. These backgrounds preserve small $\mathcal{N}=(4,4)$ superconformal symmetry and are conjectured to be dual to the symmetric orbifold CFT  \cite{Giveon:1998ns, deBoer:1998kjm, Lunin:2000yv, Maldacena:1999bp, Giveon:1999zm, Gaberdiel:2010pz, Gukov:2004ym}. A wealth of results supports this duality, ranging from precise checks of the AdS/CFT correspondence to explicit worldsheet-level quantisations of string theory in these geometries \cite{Couzens:2021veb,Haghighat:2015ega,Couzens:2019wls}.

More broadly, AdS$_3$ vacua play a central role in numerous physical contexts. They frequently arise as near-horizon geometries of black strings, providing an ideal setting for counting microstates via the Bekenstein-Hawking entropy formula, as initiated in the seminal work of Strominger and Vafa \cite{Strominger:1996sh}. Furthermore, AdS$_3$ backgrounds have been instrumental in the holographic realisation of $c$-extremization \cite{Benini:2013cda,Bah:2019jts,Couzens:2018wnk,Couzens:2022agr}, as well as in the geometric description of surface defects within higher-dimensional CFTs \cite{DHoker:2008rje,DHoker:2009lky,Faedo:2020nol,Lozano:2022ouq,Anabalon:2022fti}.

Given this broad spectrum of applications, the classification of supersymmetric AdS$_3$ solutions has been the focus of sustained interest \cite{Macpherson:2023cbl, Tong:2014yna, Lozano:2015cra, Lozano:2015bra, Kelekci:2016uqv, Couzens:2017way, Eberhardt:2017pty, Dibitetto:2017tve, Dibitetto:2017klx, Datta:2017ert, Couzens:2017nnr, Gaberdiel:2018rqv, Eberhardt:2018sce, Dibitetto:2018ftj, Dibitetto:2018iar, Eberhardt:2018ouy, Macpherson:2018mif, Lozano:2019emq, Lozano:2019jza, Lozano:2019zvg, Lozano:2019ywa, Passias:2019rga, Eberhardt:2019ywk, Couzens:2019iog, Couzens:2019mkh, Legramandi:2019xqd, Lozano:2020bxo, Faedo:2020nol, Dibitetto:2020bsh, Passias:2020ubv, Faedo:2020lyw, Legramandi:2020txf, Couzens:2021tnv, Couzens:2021veb, Macpherson:2021lbr, Macpherson:2022sbs, Conti:2024rqy, Conti:2024rwd}. Unlike their higher-dimensional counterparts, AdS$_3$ solutions admit a wide array of superconformal algebras and accommodate diverse amounts of preserved supersymmetry, making their classification both challenging and conceptually rich.

In particular, recent efforts have focused on AdS$_3$ backgrounds preserving small $\mathcal{N}=(4,0)$ and $\mathcal{N}=(6,0)$ supersymmetry. A number of new AdS$_3$ solutions with $\mathcal{N}=(4,0)$ supersymmetry have been constructed in ten- and eleven-dimensional supergravity, expanding the landscape of known vacua and offering novel insights into their dual two-dimensional conformal field theories \cite{Lozano:2019ywa, Lozano:2019zvg, Lozano:2019jza, Lozano:2019emq, Lozano:2020bxo}. More recently, a new class of AdS$_3$ solutions preserving $\mathcal{N}=(6,0)$ supersymmetry has been constructed in type IIA supergravity, opening a promising avenue for exploring holography in less conventional supersymmetric settings \cite{Macpherson:2023cbl, Lozano:2024idt}. These developments not only broaden our understanding of the AdS$_3$/CFT$_2$ correspondence but also provide new opportunities to study dual CFTs with exotic amounts of supersymmetry and R-symmetry structures. In this context, it is natural to investigate how such solutions respond to controlled geometric deformations. Among these, TsT transformations---a sequence of T-duality, coordinate shift, and a second T-duality---offer a simple yet powerful tool for generating new supersymmetric backgrounds with modified global symmetries \cite{Lunin:2005jy}, potentially enriching the holographic dictionary and unveiling novel classes of dual field theories. In this work we will use this technique to construcut new solutions in IIA supergravity.

To analyse the supersymmetry of the new backgrounds, we employ the mathematical framework of G-structures, which provides a systematic approach to characterising preserved Killing spinors and the geometric constraints they impose. One of the central goals in string and M-theory has been the systematic classification of flux backgrounds that preserve a given amount of supersymmetry. While solving the Killing spinor equations directly is often technically demanding, G-structures offer a powerful geometric reformulation of the problem. The existence of globally defined spinors on a manifold implies a reduction of its structure group to a subgroup $G \subset SO(d)$, equipping the manifold with a $G$-structure. This structure is encoded in a set of differential forms—constructed as spinor bilinears—that obey algebraic and differential conditions dictated by supersymmetry. This perspective transforms the task of classifying supersymmetric solutions into the analysis of differential systems defined by these forms, seminal examples include \cite{Gauntlett:2002sc,Gauntlett:2003cy,Gauntlett:2004zh}.

This geometric approach was further refined in the context of type II supergravity through the pure spinor formalism \cite{Grana:2005sn,Grana:2006kf}, which recasts the supersymmetry constraints as differential conditions on a pair of compatible polyforms $\Psi_{\pm}$. These polyforms define a generalised G-structure and encode the preserved supersymmetries of the background. The intrinsic torsion associated with the $G$-structure—captured by the non-closure of the polyforms—encodes the backreaction of fluxes and sources on the geometry. This formalism provides a unified and elegant framework for describing both geometric and non-geometric flux compactifications \cite{Shelton:2005cf,Andriot:2011uh}, and has also been shown to be effective in the construction and classification of broad classes of supersymmetric solutions.

With the advent of the supersymmetry conditions developed in \cite{Dibitetto:2018ftj}, these methods have proven particularly well suited to the study of AdS$_3$ backgrounds, where they not only enable the systematic construction of large families of solutions but also facilitate the analysis of deformations. These features play a central role in the present work.

The paper is organised as follows. In section~\ref{sec2}, we review the  AdS$_3$ solutions constructed in~\cite{Macpherson:2023cbl, Lozano:2019emq}, which serve as seed solutions for our deformations. 
We provide an overview and expand upon the existing discussion of their geometric structure, the associated G-structure bilinears, and the supersymmetry-preserving conditions they satisfy under defomrations. Moreover, supersymmetric probe 
branes are studied for the backgrounds in \cite{ Lozano:2019emq} using calibrations. 
Section~\ref{sec:deformedSolutions} introduces three new families of backgrounds obtained by performing TsT transformations along the internal space of the solutions. Each family is analysed separately: in section~\ref{sec:solIdentity} we study the massive IIA backgrounds while in section~\ref{masslessII} we turn to massless IIA solutions, both of them preserving small $\mathcal{N}=(4,0)$ supersymmetry. Section~\ref{sol20} presents a new class of $\mathcal{N}=(2,0)$ solutions. We provide full expressions for the fluxes, metric, and dilaton in each case, and discuss their isometries.
Section~\ref{sec:defBi} is devoted to the analysis of the G-structures supported by the deformed solutions. We use the spinor bilinear formalism to track the transformation of the geometric structure under TsT deformations and to characterise the residual supersymmetry. This section illustrates in detail the transitions between SU(3), SU(2), and identity structures on the internal manifolds. In subsection~\ref{susybranesdef}, we study the supersymmetric D-brane embeddings compatible with the deformed geometries. We compute the pullbacks of the RR potentials and show how the calibrated embeddings persist or are modified after the TsT transformations. We also analyse the emergence of new brane sources induced by the deformation.
Finally, in section~\ref{conclusions} we summarize our main findings and outline several directions for future work. Several appendices complement the main text. Appendix~\ref{sec:AppexA} contains the details of the TsT transformation on polyform bilinears, Appendix~\ref{apppage} provides technical computations relevant for the Page charges and D-brane sources, and Appendix~\ref{app:redCP3} presents the construction of Killing spinors on $\mathbb{CP}^3$.

\section{AdS$_3$ solutions in massive IIA}\label{sec2}
In this section we briefly review some of the features of the solutions classified in  \cite{Lozano:2019emq} and \cite{ Macpherson:2023cbl}.
The solutions are in massive IIA and preserve small $\mathcal{N}=(4,0)$ and $\mathcal{N}=(6,0)$ supersymmetry respectively. We will mostly focus on the 
 supersymmetry and G-structures supported by these solutions using bispinor techniques, as this will be useful for characterising the new solutions obtained by
 deformations in terms of their preserved supersymmetries and associated G-structures. Along the way, we will introduce new expressions and additional inputs 
 that contribute to the discussion of these solutions in the context of this work.

\subsection{AdS$_3$ solutions with small $\mathcal{N}=(4,0)$ supersymmetry}\label{solution1}
We will start with the solution presented in \cite{Lozano:2019emq}. These solutions are of the warped form AdS$_3\times$S$^2$
 and there exist two classes when we impose the remaining space, not necessarily endowed with isometries, supports an SU(2)-structure. 
Here we will focus on a subclass of class I solutions for which the  NS sector is explicitly given by
\begin{equation}
  \begin{split}\label{back1}
& ds^2= \frac{u}{\sqrt{h_4 h_8}}\bigg(ds^2(\text{AdS}_3)+\frac{h_8 h_4 }{g_1}ds^2(\text{S}^2)\bigg)+\frac{\sqrt{h_4 h_8}}{u} d\rho^2+\sqrt{\frac{h_4}{h_8}}ds^{2}(\text{CY}_2),\\ 
e^{\Phi}=& \frac{2h_4^{1/4}}{h_8^{3/4}}\sqrt{\frac{u}{g_1}}, \qquad  B_2= (g_2+2\pi k )\, \text{vol}( \text{S}^2),\quad  g_1=4h_4 h_8+u'^2,\quad  g_2=-\frac{1}{2}\left(\rho-\frac{u u'}{g_1}\right)
\end{split}
\end{equation}
where  $u, h_4,h_8$ are functions of $\rho$ and a large gauge transformation on the $B_2$ field with gauge parameter $k$ has been considered. The solution above is supported by the following non-trivial RR fluxes
 \begin{equation}
  \begin{split}\label{back1fluxes}
F_0&=h_8',\;\;\;\;
F_2=-\frac{1}{2}\left(h_8-\frac{h'_8 u u'}{g_1}\right) \text{vol}(\text{S}^2),\quad F_8=\star_{10} F_2,\quad F_{10}=-\star_{10} F_0, \\ 
F_4&=- \bigg(\left(\frac{u u'}{2 h_4}\right)'+2 h_8\bigg)  d\rho \wedge \text{vol}(\text{AdS}_3)
- h'_4\, \text{vol}(\text{CY}_2), \quad  F_6=-\star_{10} F_4,
\end{split}
\end{equation}
where from now on we will consider CY$_2=$T$^4$ and thus vol(T$^4$)=$dz_7\wedge dz_8\wedge dz_9\wedge dz_{10}$. This solution realises the $\mathfrak{su}(1,1\vert 2)/\mathfrak{u}(1)$ superconformal algebra which is characterised by the R-symmetry SU(2)$_R $, realised geometrically by the S$^2$ factor in the internal space. This background is solution to massive IIA supergravity if the functions 
 $h_{4}$ and $h_8$ are linear -in the absence of sources- whilst $u''=0$ everywhere. 
 
Following conventional holography, for which fluxes are turn into D-branes wrapping orthogonal cycles, we can understand the brane configuration that supports this solution. 
The charges of those branes are obtained by integrating the magnetic components of the Page fluxes over compact cycles. Locally, there exist solutions where the $\rho$-interval 
has compact domain and is bounded by physical singularities corresponding to Dp brane and O plane behaviours which "cap-off" the space-time \cite{Lozano:2019emq}. On more general grounds, global solutions 
can be constructed by allowing the functions $h_4, h_8$ to be piece-wise linear. This is achieved by adding D4 and D8 sources in the interior of the interval such that they allow for flux jumps 
as we cross the corresponding source. In this vein,  the $\rho$ interval is partitioned into sub-intervals at the boundary of which we have a change in slope of the linear functions determining the loci of the sources. 
In doing so we ought to impose that the piece-wise linear functions are convex as well as  ensure the continuity of the NS fields of the solution across the intervals. 
For $\rho \in [\rho_i, \rho_f]$ the following conditions are imposed to the functions determining the solutions

 \begin{equation}\label{conditions1}
\begin{array}{@{}l@{\quad}l@{}}
  \text{Piece-wise linear profile} & 
  h_{4,8}(\rho) = h^{(k)}_{4,8}\Theta[k, k+1], \\[1ex]
  
  \text{Boundary conditions} & 
  h_{4,8}(\rho)\vert_{\rho=\rho_i, \rho_f} = 0, \\[1ex]
  
  \multirow{3}{*}{\parbox{4cm}{\raggedright Continuity conditions}} &
  h^{(k-1)}_{4,8}(\rho)\vert_{\rho= k} = h^{(k)}_{4,8}(\rho)\vert_{\rho= k}, \\[1ex]
  
  & u^{(k-1)}(\rho)\vert_{\rho= k} = u^{(k)}(\rho)\vert_{\rho= k}, \\[1ex]
  
  & u'^{(k-1)}(\rho)\vert_{\rho= k} = u'^{(k)}(\rho)\vert_{\rho= k}
\end{array}
\end{equation}
where $h^{(k)}_{4,8}$ are linear functions in the $[k, k+1]$ interval\footnote{we note that the boundary conditions at the end of the interval need not hold for both functions in order to achieve an interval with compact support. See \cite{Filippas:2020qku} for details.} and, depending on the D-brane source we are looking at, we allow jumps on either $h'^{(k)}_{4}$, for a D4 source, or $h'^{(k)}_{8}$, for a D8 source.
 By explicitly working the quantisation of the page fluxes (see Appendix \ref{apppage}) we obtain a D2-D4-NS5-D6-D8 brane intersection in
 $\text{Mink}_2\times \mathbb{R}^{3}\times \text{T}^4\times I_{\rho}$ depicted in Table \ref{branesetup} where $\mathbb{R}^3\rightarrow (r,S^2)$,
  realises the $S^2$ associated to the R symmetry of the solutions.  
 In the brane intersection, the D2 and D6 branes generate the AdS$_3$ near horizon geometry whilst D4 and D8 are flavour branes backreacting it. 
 The latter are wrapping AdS$_3\times$S$^2$ and must correspond to supersymmetric embeddings whilst the former, since the solution is supersymmetric,  need only be stable.
 We will check this explicitly in the next subsection for comparison in latter sections with the results associated to the new solutions. 
 \begin{table}[h!]
\centering
\begin{tabular}{c|c c|lcclccc|c c }
  & $t$ & $x$ & $r$ & $\theta$ & $\phi$ & $z_7$ & $z_8$ & $z_9$ & $z_{10}$ & $\rho$ \\ [0.5ex] 
 \hline\hline 
 D8 & $\bullet$ & $\bullet$ & $\bullet$ & $\bullet$ & $\bullet$ & $\bullet$ & $\bullet$ & $\bullet$ & $\bullet$ & $\cdot$ \\ 
 \hline
 D6 & $\bullet$ & $\bullet$ & $\cdot$ & $\cdot$ & $\cdot$ & $\bullet$ & $\bullet$ & $\bullet$ & $\bullet$ & $\bullet$ \\ 
 \hline 
  D4 & $\bullet$ & $\bullet$ & $\bullet$ & $\bullet$ & $\bullet$ & $\cdot$ & $\cdot$ & $\cdot$ & $\cdot$ & $\cdot$ \\ 
 \hline
  D2 & $\bullet$ & $\bullet$ & $\cdot$ & $\cdot$ & $\cdot$ & $\cdot$ & $\cdot$ & $\cdot$ & $\cdot$ & $\bullet$ \\
  \hline\hline
  NS5 & $\bullet$ & $\bullet$ & $\cdot$ & $\cdot$ & $\cdot$ & $\bullet$ & $\bullet$ & $\bullet$ & $\bullet$ & $\cdot$ 
\end{tabular}
\caption{The D2-D4-NS5-D6-D8 brane configuration associated to the AdS$_3$ solution. The coordinates $(t,x)$ parametrise Mink$_2$, $(\theta,\phi)$ the S$^2$ and $z_i$s the T$^4$ directions. }
\label{branesetup}
\end{table}

\subsubsection{Supersymmetry and G structures}\label{Gstr1}
In the previous subsection, we summarised the results of \cite{Macpherson:2023cbl}, which show that the background defined by equations (\ref{back1})-(\ref{back1fluxes})  preserve small $\mathcal{N}=(4,0)$ supersymmetry
and support an SU(2) structure in the internal space. The purpose of this subsection is to show how this is realised explicitly. Alongside, we will discuss the way 
supersymmety can be preserved after certain transformations are applied to the solution.

The family of solutions given in equations (\ref{back1})-(\ref{back1fluxes}) were constructed using Killing spinor techniques, where supersymmetry is reformulated in terms of equations 
that involve forms and exterior algebra. In order to proceed, we need a type II supergravity background preserving the isometries of AdS$_3$.
The more general ansatz is of the form 
\begin{equation}
\begin{split}
	ds^2=&e^{2A} ds^2(\text{AdS}_3)+ds^2({M_7}), \quad F=f+e^{3A}\text{vol}(\text{AdS}_3)\wedge\star _ {7}\lambda(f), \\ 
\end{split}
\end{equation}   
where $f$ are the magnetic components of the RR polyform and $\lambda(X_p)=(-1)^{[p/2]}X_p$. The dilaton $\Phi$, warping factor $e^A$ 
as well as the magnetic NS flux $H$ have only support on the internal space M$_7$. 
Supersymmetric solutions of this form imply the existence of a pair of Majorana-Weyl
spinors $\epsilon_{1,2}$ that factorise into AdS$_3$ and M$_7$ components,
\begin{equation}
\epsilon_1=\sum^4_{I=1}\xi^I\otimes \left(\begin{smallmatrix} 1 \\ 0 \end{smallmatrix}\right)\otimes \chi^I_1, \quad \epsilon_2=\sum^4_{I=1}\xi^I\otimes \left(\begin{smallmatrix} 0 \\ 1 \end{smallmatrix}\right)\otimes \chi^I_2,
\end{equation}
where $\xi$ and $\chi_{1,2}$ are Majorana Killing spinor on AdS$_3$ and M$_7$ respectively.

The method based on G-structures boils down to constructing spinors on M$_7$ transforming 
in the correct representation of the R-symmetry group necessary to realise $\mathcal{N}=(p,q)$ supersymmetry in general. In the case at hand, for small $\mathcal{N}=(4,0)$
the R-symmetry is SU(2)$_R$ and the spinors must transform under the $\mathbf{2}\oplus\bar{\mathbf{2}}$ representation. This imposes constraints on M$_7$ since
it must support this isometry and spinors transforming in the aforementioned representation. The easiest way to ensure this is to assume there is a round S$^2$. 
The metric of the internal space then decomposes as 
\begin{equation}
	ds^2=e^{2A} ds(\text{AdS}_3)+e^{2C}ds^2(\text{S}^2)+ds^2({M_5}),
\end{equation}   
where the functions $(e^A,e^C)$ can have dependence on $M_5$ only. 

For AdS$_3$ solutions in IIA supergravity to preserve $\mathcal{N}=(1,0)$ supersymmetry, the following conditions were derived
in \cite{Dibitetto:2018ftj,Passias:2019rga,Macpherson:2021lbr},
\begin{align}
\label{eq:BPSequations}
&d_H(e^{A-\Phi}\Psi_{-})=0,\nonumber\\
&d_H(e^{2A-\Phi}\Psi_+)- 2me^{A-\Phi}\Psi_{-}=\frac{e^{3A}}{8}\star{}_7\lambda(f),\nonumber\\
&(\Psi_-,f)_7=\mp\frac{m}{2}e^{-\Phi}\text{vol}(M_7),
\end{align}
where  $(X,Y)_7=X\wedge\lambda(Y)|_7$ is the 7d Mukai Pairing, $\text{vol}({M_7})$ is the volume on $M_7$. 
The previous expressions are defined in terms of two bi-spinors
\begin{equation}
	\label{eq:bilinear}
	\Psi_{+}+i\Psi_{-}
	=\frac{1}{8}\sum_{j=0}^{d=7}\frac{1}{j!}\;(\chi_2^{\dag}\gamma_{a_1,...,a_j}\chi_1)\;e^{a_1}\wedge...\wedge e^{a_j},
\end{equation}
which are defined in terms of the internal spinors $\chi_{1,2}$ obeying an equal spinor norm condition, $|\chi_1|^2=|\chi_2|^2=e^A$. 

Moreover,
since the internal space factorises to accommodate for an S$^2$ so does the internal spinors. Explicitly they are
\begin{align}\label{eq:bi12}
&\chi_1^{I}=\frac{1}{2}e^{\frac{A}{2}}\mathcal{M}^{I}_{\delta\sigma}\bigg(\left(\cos\frac{\beta}{2}+\sin\frac{\beta}{2}\right)\xi^{\sigma}+\left(\cos\frac{\beta}{2}-\sin\frac{\beta}{2}\right)\hat{\xi}^{\delta}\bigg)\otimes\eta^{\sigma}_{1},\nonumber\\
&\chi_2^{I}=\frac{1}{2}e^{\frac{A}{2}}\mathcal{M}^{I}_{\delta\sigma}\bigg(\left(\cos\frac{\beta}{2}-\sin\frac{\beta}{2}\right)\xi^{\delta}+\left(\cos\frac{\beta}{2}+\sin\frac{\beta}{2}\right)\hat{\xi}^{\delta}\bigg)\otimes\eta^{\sigma}_{2},
\end{align} 
where $\beta$ is a function on M$_5$,  $\xi^{\delta}$, $\hat{\xi}^{\delta}$ are independent SU(2) doublets of Killing spinors on S$^2$, $\eta_1^{\sigma}$, $\eta_2^{\sigma}$ spinor doublets on $M_5$ and $\mathcal{M}^{I}=(\sigma_2\sigma_1,\sigma_2\sigma_2,\sigma_2\sigma_3,-i\sigma_2)^I$, for $I=1,2,3,4$.

Moreover, the spinors $\eta_1$ and $\eta_2$ can define either an SU(2) or an identitty structure in five dimensions.  To be more concrete, for $\eta_1=\eta$,  we
can write 
\begin{equation}
\eta_2= a_1 \eta +a_2\eta^c +\frac{a_3}{2}\overline{w}\eta, \quad   \vert \eta \vert=1 , 
\end{equation}
where $a_1, a_2\in \mathbb{C}$ and $a_3\in \mathbb{R}$ are in general point-dependent functions and $w$ is a complex one-form. The most generic bispinor constructed from these
spinors defines an identity structure whilst if $a_3=0$ an SU(2) structure. 
For class I solutions, the five-dimensional spinors are parallel, namely $a_3=0$.  One can also consistently choose  $a_2=0$
witht he remaining point-dependent function in M$_5$ trivial with M$_4$ a conformally Calabi-Yau two-fold \cite{Lozano:2019emq}.
The five-dimensional bispinors parametrising the structure are \cite{Apruzzi:2015zna}
\begin{align}\label{eq:bilinearsSU(2)}
	\psi_+^1&=(\eta\otimes\eta^{\dag})_+=\frac{1}{4}e^{-ij},\qquad\qquad \psi_+^2=(\eta\otimes\eta^{c\dag})_+=\frac{1}{4}\omega,\nonumber\\
	\psi_-^1&=(\eta\otimes\eta^{\dag})_-=\frac{1}{4}v\wedge e^{-ij},\qquad \psi_-^2=(\eta\otimes\eta^{c\dag})_-=\frac{1}{4}v\wedge \omega,
\end{align}
which are given in terms of a real two-form $j$ and a holomorphic two-form $\omega$ which are canonical in CY$_2$,
\begin{align}\label{su2forms}
j=e^7\wedge e^8+e^9\wedge e^{10},\qquad\qquad \omega=(e^7+i e^8)\wedge(e^9+i e^{10}), 
\end{align}
and a real one-form $v$ defining a transverse foliation \footnote{We are numerating the vielbeins as: $(e^1,e^2,e^3)\to $AdS$_3$,  $(e^4,e^5)\to$S$^2$, $e^6\to \rho$ and $(e^7,e^8,e^9,e^{10})\to$ M$^4$.}. They all satisfy
\begin{align}
\label{eq:SU(2)onM4}	j\wedge\omega=\omega\wedge\omega=0,&\qquad \omega\wedge \bar{\omega}=2j\wedge j=\text{vol}(M_4),\\ 
	\iota_v v=1,&\qquad \iota_v j=\iota_v \omega=0, 
\end{align} 
 defining the SU(2)-structure.  For instance, if we choose CY$_2=$ T$^4$, 
 a natural set of vielbeins defining the structure is given by
  \begin{equation}
 \begin{split}
 \qquad e^{\rho}=  \frac{(h_8 h_4)^{1/4}}{\sqrt{u}}  d\rho, \quad e^{i}=\left(\frac{h_4}{h_8}\right)^{1/4} dz_i, \quad i=7, \ldots, 10, 
 \end{split}
 \end{equation}
which can be easily shown to satisfy (\ref{eq:SU(2)onM4}) for $v=e^{\rho}$.

 Finally, we can express the seven-dimensional bi-spinors in terms of the SU(2) structure as follows \cite{Lozano:2019emq, Macpherson:2022sbs} 
 \begin{align}
&\Psi_+^{(2)}=\frac{1}{2}\; \text{Re}\left[\psi_+^1\wedge\left((1+i e^{2C}y_3\text{vol}(\text{S}^2))\cos\beta+i(y_3+ie^{2C}\text{vol}(\text{S}^2))\sin\beta\right)+i\psi_-^1\wedge K_3 e^C\right.\nonumber\\
&\qquad\qquad\quad\left.+\psi_+^2\wedge(e^{2C}\text{vol}(\text{S}^2))\cos\beta+\sin\beta) (i y_1+y_2)e^{2C}+\psi_-^2\wedge(K_2+iK_1)e^{C}\right],\nonumber\\
&\Psi_-^{(2)}=-\frac{1}{2}\; \text{Im}\left[\psi_+^1\wedge dy_3 e^{C}-\psi_-^1\wedge\left((y_3+i e^{2C}\text{vol}(\text{S}^2))\cos\beta+(i-y_3e^{2C}\text{vol}(\text{S}^2))\sin\beta\right)\right.\nonumber\\
&\qquad\qquad\quad\left.+\psi_+^2\wedge (dy_1-idy_2)e^{C}-\psi_-^2\wedge(\cos\beta-e^{2C}\text{vol}(\text{S}^2))\sin\beta)(y_1-iy_2)\right],
 \end{align}
where $y_i, \, i=1,2,3$ are embedding coordinates for the S$^2$, $K_i$ the dual forms to the Killing vectors $K^i$ and we have use the 
upper index in the bispinors to indicate we used the representative $I=2$ in (\ref{eq:bi12}) to construct them. 
These bi-spinors solve the supersymmetric system (\ref{eq:BPSequations}) for the supergravity fields in equations (\ref{back1})-(\ref{back1fluxes}) realising $\mathcal{N}=(1,0)$. We generate the other three sectors
by means of an SU(2)$_R$ rotation as long as the fields of the solution are singlets under SU(2)$_R$, giving three independent bispinors which solve the 
supersymmetric system for the supergravity fields (\ref{back1})-(\ref{back1fluxes}) for a total of $\mathcal{N}=(4,0)$. 

In this work we will construct new solutions to massive and massless IIA supergravity by applying  a series  of transformations involving dualities along M$_5$. In doing so
we ensure supersymmetry will be fully preserved since the five-dimensional spinor is not charged under SU(2)$_R$. These transformations though, will 
generalise the G-structure defined by the seed solution as we will discuss in latter sections.

\subsubsection{Supersymmetric D-branes} \label{susybranes}
 In this section we will explicitly check that the D-branes in the supersymmetric configuration of Table \ref{branesetup} giving rise to the AdS$_3$ solution are of minimal energy, that is to say that they feel no force. 
 The world-volume action for a D-brane contains two contributions given by the DBI and WZ actions. Explicitly 
 \begin{equation}
S_{\text{DBI}}=-T_p \int e^{-\Phi}\sqrt{\mid {\cal P}[g+\mathcal{F}]\mid }dX^{p+1},\quad  S_{\text{WZ}}=\pm T_p \int {\cal P} [C\wedge e^{ \mathcal{F}}+F_0 \sigma],
\end{equation}
where $d\sigma=e^{\mathcal{F}}$ and  $\cal{P}$ denotes the pull-back to the world-volume of the brane with coordinates $X^{\mu}$ and $\mathcal{F}=B_2+f_2$ with $f_2$ the worldvolume flux on the brane
and $\pm$ indicating if we have branes or anti-Dp branes. 
A D-brane 
of minimal energy satisfies $S_{\text{DBI}}+S_{\text{WZ}}=0$ and in what follows we will prove the D-branes in the solution satisfy this condition.  

Throughout this section we will use the unit radius AdS$_3$ metric in Poincare coordinates. Namely
\begin{equation}
ds^2(\text{AdS}_3)=r^2(-dt^2+dx^2)+\frac{dr^2}{r^2}. 
\end{equation}
The C-potentials associated to the fluxes in equation (\ref{back1fluxes}), in a convenient gauge,  are given by
\begin{equation}
\begin{split}
\label{potentials}
C_{1}=&\frac{1}{2}(h_8-(\rho- 2\pi k)h'_{8})\, \eta , \quad\quad  d\eta=-\text{vol}(\text{S}^2), \\
C_3=&\frac{1}{2 h_4^2}dt\wedge d\rho\wedge \left((h_4 g_1-u u' h_4')\, r \,x\, dr+ \sqrt{h_4^3  h_8 g_1}\,d( r^2\, x)\right)-z_7 h'_4\text{vol}(\text{T}^{3}), \\
C_5=&-\frac{u^2}{4 h_4}\text{vol}(\text{AdS}_3)\wedge \text{vol}(\text{S}^2)+\frac{1}{2}(h_4-\rho h'_{4}) \text{vol}(\text{T}^4)\wedge \eta  \\
&-\left(\frac{2 h_4 u' u g_1-h_4'u^2(g_1+4h_4 h_8)}{8 h_4^2 g_1}r^2 dt\wedge dx\wedge d\rho- g_2 z_7 h_4'\,\text{vol}(\text{T}^3)\right)\wedge \text{vol}(\text{S}^2),\\
C_7=&\frac{h_4}{h_8}\left(C_3+\frac{2 g_1 h_4 h_8 -u u' \partial_{\rho}(h_4 h_8)}{2 h_4^2} r \, x dt\wedge dr\wedge d\rho \right)\wedge \text{vol}(\text{T}^4),\\
C_9=&\frac{h_4}{h_8}\left(C_5-\frac{2 f_1 h'_4 h_8 - u'^2 \partial_{\rho}(h_4 h_8)}{8 h_4^2 h_8 f_1} u^2\, r^2  dt\wedge dx\wedge d\rho \right)\wedge \text{vol}(\text{S}^2) \wedge \text{vol}(\text{T}^4),
\end{split}
\end{equation} 
where they all satisfy $F=d_{H}C+F_0 e^{B_2}$ where $F, \, C$ are the RR  field strength and potential polyforms respectively\footnote{We have explicitly $F=F_0+F_2+F_4+F_6+F_8+F_{10}$ and $C=C_1+C_3+C_5+C_7+C_9$ } and $d_H=d-H\wedge$ with  $H=dB_2$.

We will start with a D2 brane wrapping the three-submanifold $\Sigma^{(3)} = (\text{Mink}_2, \rho)$. The DBI Lagrangian density for
this brane is 
\begin{equation}
e^{-\Phi}\sqrt{\mid {\cal P}[g+B_2]\mid }\Big\vert_{\Sigma^{(3)}}=\frac{1}{2}r^2 \sqrt{\frac{h_8}{h_4} g_1},
\end{equation}
The WZ term using the C-potentials in equation (\ref{potentials}) give
\begin{equation}
C_3-C_1\wedge B_2\Big\vert _{\Sigma^{(3)}}=\frac{1}{2}r^2 \sqrt{\frac{h_8}{h_4} g_1},
\end{equation}
showing that the D2 brane is BPS. 

We consider now D6 brane with worldvolume spanning $\Sigma^{(7)} = (\text{Mink}_2, \rho, \text{T}^4)$. 
The DBI Lagrangian density and WZ terms for this brane read
\begin{equation}
\begin{split}
e^{-\Phi}\sqrt{\mid {\cal P}[g+B]\mid }\Big\vert_{\Sigma^{(7)}}=&\frac{1}{2}r^2 \sqrt{\frac{h_4}{h_8} g_1},\\
C_7-C_5\wedge B_2 \Big\vert_{\Sigma^{(7)}}=&-\frac{1}{2}r^2 \sqrt{\frac{h_4}{h_8} g_1},
\end{split}
\end{equation}
which means that an anti-D6 brane is stable. As we pointed out above, the D2-D6 branes correspond to colour branes.
Their backreaction give rise to the near-horizon AdS$_3$ geometry and as such their number in each interval are associated to the ranks 
of gauge groups for the dual field theory \cite{Lozano:2019zvg}. 
Notice that this picture arises since we are considering large gauge transformation on the $B_2$. If this were not the case, a worldvolume flux in the D8 brane will be needed.
The D6 would be part of a D8-D6 bound state, namely there would be induced D8 charge in the D6 due to the non-zero worldvolume flux.

For the D4 branes lying along $\Sigma^{(5)} = (\text{AdS}_3, S^2)$ we find that the DBI and WZ terms 
give 
\begin{equation}\label{dd4}
\begin{split}
e^{-\Phi}\sqrt{\mid {\cal P}[g+B]\mid }\Big\vert_{\Sigma^{(5)}}=&\frac{1}{4}\frac{u \sqrt{4 h_4 h_8 u^2+(u u'-g_1 (\rho-2\pi k))^2} r \sin\theta }{ h_4 \sqrt{g_1}},\\
C_5-C_3\wedge B_2\Big\vert_{\Sigma^{(5)}}=&-\frac{1}{4}\frac{u^2}{h_4}r \sin \theta. 
\end{split}
\end{equation}
We see that in order for the (anti) D4 to be stable we need to placed them at special points, $\rho=2\pi k$ for $k\in \mathbb{Z}$,  along the interval as a consequence of the large gauge transformation on the B$_2$ field.

Lets us finally consider the D8 brane with worldvolume spanning $\Sigma^{(9)} = (\text{AdS}_3, S^2, \text{T}^4)$. 
The contributions from the Lagrangian densities read
\begin{equation}\label{dd8}
\begin{split}
e^{-\Phi}\sqrt{\mid {\cal P}[g+B]\mid }\Big\vert_{\Sigma^{(9)}}=&\frac{1}{4}\frac{u \sqrt{4 h_4 h_8 u^2+(u u'-g_1 (\rho-2\pi k))^2} r \sin \theta }{ h_8 \sqrt{g_1}},\\
C_9-C_7\wedge B_2\Big\vert_{\Sigma^{(9)}}=&-\frac{1}{4}\frac{u^2}{h_8}r \sin \theta ,
\end{split}
\end{equation}
In the same way as for the D4 brane, we see that stability requires to place the place the D8 branes at $\rho=2\pi k$.  

Notice that since the D4 and D8 branes correspond to flavour branes (see Appendix \ref{apppage}) the stability condition does not guarantee that the embedding of the corresponding brane is supersymmetric.
Namely, we ought to check the $\kappa$-symmetry constraint for the embeddings.  However, the bispinors constructed in section (\ref{Gstr1}) can be used to determine the supersymmetric brane probes allowed by the solution. 
 This is done in terms of calibrations. Namely, for a Dp brane with worldvolume along AdS$_3$ and  wrapping a internal submanifold we require 
  \begin{equation}\label{cal}
 e^{-\Phi}\sqrt{\mid {\cal P}[g+\mathcal{F}]\mid }\Big\vert_{\Sigma}=\pm \Psi^{(\text{cal})}_{Dp}\Big\vert_{\Sigma}=\pm 8e^{3 A-\Phi} \text{vol}(\text{AdS}_3)\wedge \Psi_{-}\wedge e^{-\mathcal{F}}\Big\vert_{\Sigma}, 
 \end{equation}
 where the pullback to the worldvolume is understood. 
 In the case at hand, the calibration forms for the D4 and D8 branes source branes with $f_2=0$ are 
\begin{align}
\Psi^{(\text{cal})}_{D4}=&\frac{u (4 h_4 h_8 u+u'(u u'-g_1(\rho-2\pi k )))}{4 h_4 g_1}\, \text{vol}(\text{AdS}_3)\wedge \text{vol} (\text{S}^2),\\
\Psi^{(\text{cal})}_{D8}=&\frac{u (4 h_4 h_8 u+u'(u u'-g_1(\rho-2\pi k )))}{4 h_8 g _1}\, \text{vol}(\text{AdS}_3)\wedge \text{vol} (\text{S}^2)\wedge \text{vol} (\text{T}^4).
\end{align}
By comparing the DBI terms for each brane from equation in (\ref{dd4}) and (\ref{dd8}) one can easily see that the corresponding sources are calibrated whenever they are placed at $\rho= 2 \pi k$ as anticipated. 
On the other hand, an easy computation shows that it is not possible to have D2 and D6 branes along AdS$_3$. 

In latter sections we will study the stability of the BPS branes above when they probe the geometries that we construct via deformations
as well as the new branes that appear as a consequence of it.


\subsection{AdS$_3$ solutions with $\mathcal{N}=(6,0)$ supersymmetry}\label{susy06}
The solutions classified in \cite{Macpherson:2023cbl}  correspond to foliations of the warped product $AdS_3\times \mathbb{CP}^3$ over an interval
supporting an SU(3) structure in the internal space. 
 The NS sector of the  solution is
\begin{align}
\label{back2}
\frac{ds^2}{2\pi}&= \frac{|h|}{\sqrt{\Delta}}ds^2(\text{AdS}_3)+\sqrt{\Delta}\bigg[ \frac{1}{4 |h|}d\rho^2+ \frac{2}{|h''|}
ds^2(\mathbb{CP}^3)\bigg],\nonumber\\[2mm]
e^{-\Phi}&=\frac{(|h''|)^{\frac{3}{2}}}{2\sqrt{\pi}\Delta^{\frac{1}{4}}},\quad B_2=c(\rho)\, J, \quad c(\rho)=4\pi \left( -(\rho-k)+\frac{h'}{h''}\right), 
\end{align}
where $h$ is a function of $\rho$, $\Delta=2 h h''-(h')^2\geq 0$, $k$ a constant which can be interpreted as performing a large transformation on $B_2$ and J is the Kahler form on $\mathbb{CP}^3$. 
For the analysis that follows, 
it will be convenient to write the $\mathbb{CP}^3$  metric as a foliation of $T^{1,1}$  over an interval, namely 
\begin{align}\label{cp3}
ds^2(\mathbb{CP}^3)&= d\xi^2+\frac{1}{4}\cos^2\xi ds^2(\text{S}^2_1)+\frac{1}{4}\sin^2\xi ds^2(\text{S}^2_2)+ \frac{1}{4}\sin^2\xi\cos^2\xi (d\psi+ \eta_1- \eta_2)^2,\nonumber\\[2mm]
&\qquad \qquad \qquad \qquad d\eta_i=-\text{vol}(\text{S}_i^2) 
\end{align}
In this parametrisation the Kahler form is given by
\begin{align}\label{cp3K}
J= -\frac{1}{4}\sin^2\xi \text{vol}(\text{S}^2_2)-\frac{1}{4}\cos^2\xi \text{vol}(\text{S}^2_1)-\frac{1}{2}\sin\xi \cos\xi d\xi\wedge (d\psi+ \eta_1-\eta_2). 
\end{align}
 For latter use, we quote the canonical choice for the frames defining the metric of the seven-dimensional internal space
\begin{equation}\label{framescp3}
\begin{split}
e^1=&2\sqrt{\pi} \frac{\Delta^{1/4}}{\sqrt{h''}} d\xi, \quad e^2=\sqrt{\pi} \frac{\Delta^{1/4}}{\sqrt{h''}} d\theta_1, \quad e^3=\sqrt{\pi} \frac{\Delta^{1/4}}{\sqrt{h''}}  \sin\theta_1 d\phi_1, \quad e^4=\sqrt{\pi} \frac{\Delta^{1/4}}{\sqrt{h''}} d\theta_2,\\
&e^5=\sqrt{\pi} \frac{\Delta^{1/4}}{\sqrt{h''}}  \sin\theta_2 d\phi_2. \quad  e^6=\sqrt{\pi} \frac{\Delta^{1/4}}{\sqrt{h''}} (d\psi+\eta_1-\eta_2), \quad e^7=\sqrt{\frac{\pi}{2}} \frac{\Delta^{1/4}}{\sqrt{h}} d\rho.
\end{split}
\end{equation}

The RR sector supporting the solution is explicitly given by
\begin{align}
F_0&=-\frac{1}{2\pi}h''',~~~~F_2=B_2 F_0+2 (h''-(\rho-k)h''')J,\nonumber\\[2mm]
F_4&=\pi \, d\left(h'+\frac{hh'h''}{\Delta}\right)\wedge\text{vol}(\text{AdS}_3)+B_2\wedge F_2-\frac12 B_2\wedge B_2 F_0\nonumber\\[2mm]
&-4\pi (2h'+(\rho-k)(-2h''+(\rho-k)h'''))J\wedge J\label{RRgeometry}.
\end{align}
where supersymmetry and Bianchi identities impose that the function $h$ is a third order polynomial, namely $h''''=0$. The solutions just described 
realise the $\mathfrak{osp}(6,2)$ superconformal algebra which is characterised by the R-symmetry group $SO(6)_R$ realised in the geometry in terms of the $\mathbb{CP}^3$ manifold. 
As was shown in \cite{Macpherson:2023cbl} when the $h$ function is a second order polynomial  (massless IIA limit) we recover -locally- the $AdS_4\times \mathbb{CP}^3$ ABJM solution \cite{Aharony:2008ug}. 
In such a limit, we use 
\begin{equation}
h=Q_{\text{D}_2}-Q_{\text{D}_4}\rho +\frac{1}{2} Q_{\text{D}_6} \rho^2,
\end{equation}
where the $Q_{\text{D}p},\, p=2,4,6$ are the quantised charges of Dp-branes in the solution. After a change of coordinates involving the coordinate $\rho$ in (\ref{back2}), the subspace (AdS$_3$,$\rho$) 
turns into AdS$_4$ (see \cite{Macpherson:2023cbl} for details). 

Local solutions within this class for $F_0\neq 0$ impose that the $r$ interval is semi-infinite. However, as we discussed earlier for the solutions in Section (\ref{solution1}), one can construct more general global 
solutions by glueing local solutions with the addition of D8 brane sources along the interval as long as we ensure the NS fields of the solution are continuous as well as the condition $-2h''+(\rho-l)h'''\geq 0$ 
satisfied at each interval. The conditions on the $h$ function are thus very similar to those quoted in (\ref{conditions1}) but for $h^{(k)}$ a third order polynomial in the $[k,k+1]$ interval and continuity for all the $h,h',h''$
across intervals whilst we allow jumps of $h'''$ at the loci of D8 brane sources. 

\subsubsection{Supersymmetry and G structures}\label{Gstr2}

In the previous section, we presented the solutions constructed in \cite{Macpherson:2023cbl}, which were also derived using Killing spinor techniques as those presented in section (\ref{Gstr1}). 
To be more precise, two classes of solutions, associated with the 
algebras $\mathfrak{osp}(6|2)$ and $\mathfrak{osp}(5|2)$, were constructed using a single parametrisation. This was achieved by expressing round $\mathbb{CP}^3$ as a fibration of $S^2$
over $S^4$, such that it provides a space that supports both the correct isometry and spinors transforming in the $\mathbf{6}$ of SO(6)$_R$.
On the other hand, when the $S^2$ fiber is pinched, the SO(6)$_R$ simmetry is broken locally to SO(5). In this case, the \textbf{6} of 
SO(6) decomposes into $\mathbf{5}\oplus \mathbf{1}$ under its SO(5) subgroup. This decomposition provides both the representation and the R-symmetry required by $\mathfrak{osp}(5|2)$.

In this work, we focus on the AdS$_3$ solutions preserving $\mathcal{N}=(6,0)$ supersymmetry.  Since it is more convenient for the analysis in latter sections, we parametrised $\mathbb{CP}^3$ in (\ref{cp3})
as a foliation of $T^{1,1}$ over an interval. In appendix \ref{app:redCP3} we deal with the technical details of constructing the Killing spinors\footnote{When we were readying our manuscript, we notice these results were presented in Appendix A of \cite{Conti:2025djz}.} on $\mathbb{CP}^3$
as a reduction of those on S$^7$, when written as a U(1) fibration over $\mathbb{CP}^3$. After such a reduction, the SO(6)$_R$  sextuplets are given by
\begin{equation}\label{eq:the66}
	\textbf{6}:~\zeta^{\cal I}_6=\frac{1}{\sqrt{2}}\left(\begin{array}{c}i(\zeta_+^1+\zeta_+^4)\\\zeta_+^1+\zeta_+^4\\i(\zeta_+^6+\zeta_+^7)\\\zeta_+^6-\zeta_+^7\\\frac{1}{\sqrt{2}}(\zeta_+^2-\zeta_+^5+(\zeta_+^3-\zeta_+^8))\\\frac{i}{\sqrt{2}}(\zeta_+^2-\zeta_+^5-(\zeta_+^3-\zeta_+^8))\end{array}\right)^{\cal I},~~~~ \hat{\zeta}^{\cal I}_6=i\gamma_7\zeta^{\cal I}_6,
\end{equation}
which are analogous to the spinors in (A.23) of \cite{Macpherson:2023cbl}. We will avoid giving the explicit form of all these spinors. 
However, below we will quote the ones that will be useful for the analysis of the next section.

Using the spinors in (\ref{eq:the66}), we can construct the seven-dimensional spinors of the solution in the following way \cite{Macpherson:2023cbl}
\begin{align}
	\chi_1^\mathcal{I}&=\cos\left(\frac{\beta_1+\beta_2}{2}\right)\; \zeta_6^\mathcal{I}+\sin\left(\frac{\beta_1+\beta_2}{2}\right)\; \hat{\zeta}_6^\mathcal{I},\nonumber\\ \chi_2^\mathcal{I}&=\cos\left(\frac{\beta_1-\beta_2}{2}\right)\; \zeta_6^\mathcal{I}+\sin\left(\frac{\beta_1-\beta_2}{2}\right)\; \hat{\zeta}_6^\mathcal{I},
\end{align}
where $\beta_1=\beta_1(\rho)$ and $\beta_2=\beta_2(\rho)$.  
Taking as representatives the $(\chi_1^5,\chi_2^5)$ spinors, namely $\mathcal{I}=5$ in (\ref{eq:the66}), the seven-dimensional bispinors (\ref{eq:bilinear}) can be written in terms of an SU(3) structure,
\begin{align}\label{eq:SU(3)bilinears}
	\Psi_{+}^{(5)}=\frac{1}{8}\;\text{Re}[e^{i\beta_2}e^{-i\mathcal{J}}-e^{i\beta_1}\Omega\wedge V],\qquad
	\Psi_{-}^{(5)}=\frac{1}{8}\;\text{Im}[e^{i\beta_2}e^{-i\mathcal{J}}\wedge V+e^{i\beta_1}\Omega],
\end{align}
where the one-form, real two-form and complex three-form  (V, $\mathcal{J}$,  $\Omega$), in terms of the seven-dimensional vielbein (\ref{framescp3}), are given by 
\begin{equation}
\begin{split}
	\mathcal{J}=&-(e^1\wedge e^6+c_{2\xi}(e^2\wedge e^3-e^4\wedge e^5)+c_{\psi} s_{2\xi}(e^2\wedge e^4+e^3\wedge e^5))\\
	&+s_{2\xi}s_\psi(e^3\wedge e^4-e^2\wedge e^5), \\
	V=&e^7,\\
	\Omega =&s_{2\xi}(e^1\wedge e^2\wedge e^3-e^1\wedge e^4\wedge e^5)-c_\psi c_{2\xi}(e^1\wedge e^2\wedge e^4+e^1\wedge e^3\wedge e^5)\\
	&+c_{2\xi}s_\psi(e^1\wedge e^2\wedge e^5-e^1\wedge e^3\wedge e^4)-s_\psi(e^2\wedge e^4\wedge e^6+e^3\wedge e^5\wedge e^6)\\
	&-c_{\psi}(e^2\wedge e^5\wedge e^6-e^3\wedge e^4\wedge e^6)+i\left[s_{2\xi}(e^4\wedge e^5\wedge e^6-e^2\wedge e^3\wedge e^6)\right.\\
	&\left.+c_\psi c_{2\xi}(e^2\wedge e^4\wedge e^6+e^3\wedge e^5\wedge e^6)+c_{2\xi}s_\psi(e^3\wedge e^4\wedge e^6-e^2\wedge e^5\wedge e^6)\right.\\
	&\left.-s_{\psi}(e^1\wedge e^2\wedge e^4+e^1\wedge e^3\wedge e^5)+c_\psi(e^1\wedge e^3\wedge e^4-e^1\wedge e^2\wedge e^5)\right]
\end{split}
\end{equation}
where we have used the notation $s_{x}=\sin x$ and $c_{x}=\cos x$. These forms satisfy the following relations associated to the SU(3) structure on M$_7$
\begin{align}
	\label{eq:SU(3)restrictionsonM7}
	\iota_V\mathcal{J}=\iota_V\Omega=0,\qquad\qquad\mathcal{J}\wedge \Omega=0, \qquad\qquad \mathcal{J}\wedge \mathcal{J}\wedge \mathcal{J}=\frac{3 i}{4}\Omega\wedge\bar{\Omega}.
\end{align} 
The bilinears in equation \eqref{eq:SU(3)bilinears} solve the supersymmetric system (\ref{eq:BPSequations}) for the fields in equations (\ref{back2}) and (\ref{RRgeometry}) implying $\mathcal{N}=(1,0)$ supersymmetry. 
By considering the remaining pairs of spinors in (\ref{eq:the66}) we generate five independent bispinors solving the supersymmetric system (\ref{eq:BPSequations}) with supergravity fields (\ref{back2}) and (\ref{RRgeometry}) for a total of $\mathcal{N}=(6,0)$
supersymmetries. 

In the next section, we will be performing a TsT transformation to the solution which involves choosing two U(1) isometries. We will focus in the case were the transformation is entirely performed along the directions of
$\mathbb{CP}^3$. Since we are interested in solutions preserving supersymmetry we must pick the spinors which are singlets under the two U(1) directions that give rise to bispinors uncharged under
such directions. 

\vspace{5pt}

In the case of the TsT transformation in the two U(1)'s inside the S$^2_i$ (i.e. the coordinates $\phi_1,\phi_2$ in the frames (\ref{framescp3})), 
the only spinors that are uncharged under $\partial_{\phi_1}$ and  $\partial_{\phi_2}$ turn out to be the 5th and 6th component of \eqref{eq:the66}. They are
\begin{align}
	\label{eq:two supercharges}
	\zeta_6^5=&\frac{e^{-i(\xi+\frac{\psi}{2})}}{4}\left(1-e^{2i\xi},\;1+e^{2i\xi},\;e^{i \psi}(1+e^{2i\xi}),\;e^{i \psi}(1-e^{2i\xi}),\;-(1+e^{2i\xi}),-(1-e^{2i\xi}),\right.\nonumber\\
	&\left.-e^{i\psi}(1-e^{2i\xi}),\;-e^{i\psi}(1+e^{2i\xi})\right)^T, \nonumber\\
	\zeta_6^6=&\frac{i\;e^{-i(\xi+\frac{\psi}{2})}}{4}\left(1-e^{2i\xi},\;1+e^{2i\xi},\;-e^{i \psi}(1+e^{2i\xi}),\;-e^{i \psi}(1-e^{2i\xi}),\;-(1+e^{2i\xi}),-(1-e^{2i\xi}),\right.\nonumber\\
	&\left.e^{i\psi}(1-e^{2i\xi}),\;e^{i\psi}(1+e^{2i\xi})\right)^T. 
\end{align}
No other choice is possible if we are to preserve a fraction of the original supersymmetry since despite the fact the remaining components  $\zeta^1_6$, $\zeta^2_6$, $\zeta^3_6$ and $\zeta^4_6$ 
are independent of $\psi$ they all depend non-trivially on $\phi_1, \phi_2$.\footnote{Other possibility will be to consider combinations of all three angles but will not consider that possibility here.}
In particular, this means that after the deformation process only the bispinors $\Psi^{(5)}_{\pm}$ and $\Psi^{(6)}_{\pm}$ survive and the deformed solution will preserve $\mathcal{N}=(2,0)$ supersymmetry, where the coordinate $\psi$ parametrises the U(1) R-symmetry.  Interestingly, the spinors $\zeta_6^5$ and $\zeta_6^6$ in \eqref{eq:two supercharges} are related by the shift $\psi \to \psi + \pi$, which is consistent with their interpretation as forming a U$(1)$ doublet.
From now on we will focus on the $\mathcal{N}=1$ subsector described by the representative $\mathcal{I}=5$. 


 \section{New supersymmetric AdS$_3$ solutions via deformations}\label{sec:deformedSolutions}
In this section we will obtain new solutions to massless and  massive IIA supergravity by applying TsT transformations in the internal space of the solutions  in equations (\ref{back1}) and (\ref{back2}). 
 The deformation consist on picking two U(1) isometries. We then perform a T-duality along one of them, followed by a shift parametrised by $\gamma$ along the other and then we T dualise back along 
 the first U(1). This procedure is guaranteed to map solutions to solutions \cite{Lunin:2005jy} but one ought to be careful in the choice of U(1)s if one is to preserve supersymmetry or at least a fraction of the original one.  
 For each of the cases we will study, we will point out the U(1)s we have at our disposal such that we obtain a supersymmetric background following the analysis of sections (\ref{Gstr1}) and (\ref{Gstr2}).

\subsection{A deformation in massive IIA}\label{sec:solIdentity}
We begin with the solutions of section (\ref{solution1}). As we discussed in section (\ref{Gstr1}), the solutions preserve $\mathcal{N}=(4,0)$ supersymmetry and full supersymmetry is preserved as long as 
the process does not involve the U(1) inside the S$^2$ since the four bispinors are charged under it. The only option is then to consider the U(1)s inside T$^4$. 
This solution was already reported in \cite{Zacarias:2021pfz}, but we will provide additional input since it will be our subject study in latter sections. This will help us to 
understand a properly characterise the solutions.

 We take the two U(1)s inside T$^4$  parametrised by shifts in the $(z_8,z_9)$ coordinates.  The deformation leaves the
 space transverse to T$^4$ invariant and we have that the NS sector is given by 
\begin{equation}
  \begin{split}\label{back1def}
&  ds^2= \frac{u}{\sqrt{h_4 h_8}}\bigg(ds^2(\text{AdS}_3)+\frac{h_8 h_4 }{g_1}ds^2(\text{S}^2)\bigg)+ \frac{\sqrt{h_4 h_8}}{u} d\rho^2 
 +\sqrt{\frac{h_4}{h_8}}\left( dz_{7,10}^2+G dz_{9,8}^2\right)\\ 
& e^{\tilde{\Phi}}= \sqrt{G}e^{\Phi},~~~~~~ \tilde{B}_2= B_2-\gamma G\, g\,  dz_9\wedge dz_8, \quad G=(1+\gamma^2 g)^{-1}, \quad g=\frac{h_4}{h_8}
\end{split}
\end{equation}
where $g$ is the determinant of T$^2$ and 
throughout tilde denotes deformed fields. 
Applying the transformation rule  (\ref{trule}) to the fluxes in equation (\ref{back1fluxes}) we obtain 
\begin{equation}\label{defRR1}
\begin{split}
\tilde{F}_0=&h_8',\;\;\;\;
\tilde{F}_2=F_2-\gamma h_4' dz_7\wedge dz_{10}-\gamma G\, g\, h'_8 dz_9\wedge dz_8,\\
\tilde{F}_4= &\, F_4+\gamma^2 h'_4\, G\, g\,  \text{vol}(\text{T}^4)+\\
&\frac{\gamma}{2} \left( G\, g\, \left(h_8-\frac{u u' h_8'}{g_1}\right)dz_9\wedge dz_8+\left(h_4-\frac{u u' h_4'}{f_1}\right)dz_7\wedge dz_{10}\right) \wedge \text{vol}(\text{S}^2),\\
\tilde{F}_6=&-\star_{10} \tilde{F}_4, \quad \tilde{F}_8=\star_{10} \tilde{F}_2, \quad \tilde{F}_{10}=-\star_{10} \tilde{F}_0.
\end{split}
\end{equation} 
This constitutes a solution of massive IIA provided $u''=0$ and $h''_8=h''_4=0$ away from localised sources 
and reduces to the undeformed one when the deformation parameter $\gamma$ vanishes.

Since we will be analysing D-brane embeddings for this solution, we write the C-potentials associated to the deformed fluxes (\ref{defRR1}). 
For simplicity we only quote the ones we will be needing latter on. They are
\begin{equation} \label{cpotdef}
\begin{split}
\tilde{C}_1=&\, C_1-\gamma h'_4 z_7 dz_{10} , \\
\tilde{C}_3=&\, C_3+ \, \tilde{C}_1 \wedge \tilde{B}_2+\frac{\gamma}{2}(h_4-\rho h'_4) dz_7\wedge dz_{10} \wedge \eta+\gamma \, k \pi z_7 dz_{10}\wedge \text{vol}( \text{S}^2),  \\
\tilde{C_5}= &\, C_5-\frac{\gamma}{2}\sqrt{\frac{h_4}{h_8}g_1} r^2 dt\wedge dx\wedge d\rho\wedge \left(dz_7\wedge dz_{10}-G dz_9\wedge dz_8\right)\\
-& \frac{\gamma}{2 h_8} r x \, dt\wedge dr\wedge d\rho \wedge \left(\left(- g_1+\frac{u u' h'_8}{h_8}+2\sqrt{h_4 h_8 g_1}\right)dz_7\wedge dz_{10}\right.\\
-&\left. G\left(g_1-\frac{u u' h'_4}{h_4}+2\sqrt{h_4 h_8 g_1}\right)dz_9\wedge dz_8\right)\\
 -&\frac{\gamma^2}{2} (h_4-\rho h'_4)G\, g \, \text{vol}(T^4)\wedge \eta +\gamma^2 G\, g\, h'_4 g_2 z_7 dz_{10}\wedge \text{vol}(T^2)\wedge \text{vol}(S^2), 
\end{split}
\end{equation}
where they all satisfy $\tilde{F}=d_{\tilde{H}} \tilde{C}+\tilde{F}_0 e^{\tilde{B}_2}$. 
 
 The quantisation of the fluxes (see Appendix \ref{apppage}) tell us that on top of the branes underlying the undeformed solution
 we have extra D4 branes wrapping Mink$_2\times I_{\rho}\times \text{T}^2$ and D6 branes wrapping AdS$_3\times$S$^2\times$T$^2$
 with charge 
 \begin{equation}\label{massless}
 Q^{(k)}_{\tilde{\text{D}}6}=\gamma Q^{(k)}_{\text{D}4},\in \mathbb{Z} \quad Q^{(k)}_{\tilde{\text{D}}4}=\gamma Q^{(k)}_{\text{D}2}, \in \mathbb{Z}
 \end{equation}
 which implies that for $\gamma=\frac{1}{n}$ the charges $Q^{(k)}_{\text{D}4}=\kappa n$, 
 $Q^{(k)}_{D2}=\ell n$ for $\kappa, \ell, n  \in \mathbb{Z}$. 
 The analysis of the Bianchi identities for the fluxes sourcing this branes tells us that the new $\tilde{\text{D}}$6 branes correspond to flavour branes
 since there is induced charge of the flavour D4 \cite{Zacarias:2021pfz}. This is a manifestation of the fact that the D4 expanded along T$^2$
 into the $\tilde{\text{D}}$6 that wraps this submanifold. In section \ref{susybranesdef} we will see that these branes are stable provided a magnetic worldvolume
 flux is turned on. The latter also required to solve the modified Bianchi identity for the $\tilde{\text{D}}$6 (see equation (\ref{bianchid6})).
 The same happens for the new D4 but since this is not a source brane, no modified Bianchi identity need to be solved.

 \subsection{A deformation in massless IIA}\label{masslessII}
We now consider a sub-class of solutions of (\ref{back1})-(\ref{back1fluxes}) for which $\partial_{\rho}$ is a Killing vector. 
In this case all warping factors of the solution become trivial.
The $B_2$ field can then be re-written in a way that makes the isometry in $\rho$ manifest.  We then consider the deformation along the two torus spanned by $(\rho,z_{10})$.
Once again this solution preserves all supersymmetries of the seed solution since the S$^2$ is an spectator for the deformation. 
Applying the procedure outlined above we obtain the solution
\begin{equation}
\begin{split}\label{seconddef}
ds^2=&\frac{c_0}{\sqrt{c_4 c_8}}\left(ds^2(\text{AdS}_3)+\frac{1}{4}\left(ds^2(\text{S}^2)+\gamma^2 \sqrt{\frac{c_4}{c_8}}G \, (d\psi+\cos\theta d\phi)^2\right)\right)\\
&+\sqrt{\frac{c_4}{c_8}}ds^2(\text{T}^3)+\frac{\sqrt{c_4 c_8}}{c_0}\, G\, d\rho^2,\\
\tilde{B}_2=&\, \frac{G}{2} \left(d\psi+\cos\theta d\phi \right)\wedge d\rho, \quad e^{\tilde{\Phi}}=\frac{\sqrt{ G \, c_0}}{(c_8^5\,  c_4)^{1/4}}, \quad G=(1+\gamma^2 g)^{-1},  \quad g=\frac{c_4}{c_0}, 
\end{split}
\end{equation} 
where $c_0, c_4, c_8$ are constants and we have defined $z_{10}=\frac{\gamma}{2} \, \psi$ and implemented a gauge transformation on $\tilde{B}_2$. 
The non-zero RR fluxes supporting the solution are given by
\begin{equation}
\begin{split}
\tilde{F}_2=&-\frac{1}{2}c_8\, \text{vol}(\text{S}^2), \quad \tilde{F}_4=2 c_8\, \text{vol}(\text{AdS}_3)\wedge d\rho+\tilde{F}_2\wedge \tilde{B}_2.\\
&\qquad \qquad  \tilde{F}_6=-\star_{10} \tilde{F}_4, \quad \tilde{F}_8=\star_{10} \tilde{F}_2.
\end{split}
\end{equation}
We then see explicitly that the solution gives rise to an SU(2)$_R\times$U(1) preserving squashed S$^3$. For non-zero values
of the constants $c_i$ and the deformation parameter, we can tune the squashing term to one such that we have a round S$^3$.
This enhancement is also reflected in the supersymmetries preserved by the solution since in this special case we have a second SU(2) and
potentially small $\mathcal{N}=(4,4)$. This can also be seen since the metric in equation (\ref{seconddef}) is the T-dual of the D1-D5 system 
when T-duality is performed along one of the T$^4$ directions.

\subsection{A deformation preserving $\mathcal{N}=(2,0)$ in massive IIA}\label{sol20}
We now turn to the solutions discussed in section (\ref{susy06}). As we have pointed out, we are only interested in solutions where the deformation is entirely applied
in the internal space of the solutions. Following the discussion in section \ref{Gstr2}, we learnt that in order to obtain a supersymmetric background we have to choose 
the two torus spanned by $(\phi_1, \phi_2)$ which lie inside the S$^2_{i}$ in the metric of equation (\ref{cp3}). In this way we ensure that $\mathcal{N}=(2,0)$ supersymmetry
is preserved. 

 Following the procedure described above, the new background is characterised by the NS sector 
 \begin{equation}
\label{back2def}
\begin{split}
\frac{ds^2}{2\pi}&= \frac{|h|}{\sqrt{\Delta}}ds^2(\text{AdS}_3)+\sqrt{\Delta}\bigg[ \frac{1}{4 |h|}d\rho^2+ \frac{2}{|h''|}
ds^2(\widetilde{\mathbb{CP}}^3)\bigg],\\
e^{\tilde{\Phi}}=&2\Delta^{1/4}(h'')^{-3/2}\sqrt{\pi G},\quad \tilde{B}_2=B_2+\gamma \left(\frac{c(\rho)^2}{16}Z_1-\frac{\pi^2\Delta}{(h'')^2}W\right),
\end{split}
\end{equation}
where 
\begin{equation}\label{definitionsH}
\begin{split}
ds^2&(\widetilde{\mathbb{CP}}^3)= d\xi^2+\frac{1}{4}\cos^2\xi d\theta_1^2+\frac{1}{4}\sin^2\xi d\theta_2^2+ \gamma^2 G\, \frac{\Delta \, \cos^4\xi \sin^4\xi \sin^2\theta_1 \sin^2\theta_2}{4 h''^2}d\psi^2\\
+&\frac{1}{4}\sin^2\xi\cos^2\xi \, G (d\psi+ \eta_1- \eta_2-\frac{\gamma}{4}c(\rho)\cos\theta_1\sin\theta_2\sin^2\xi d\theta_2-\frac{\gamma}{4}c(\rho)\cos\theta_2\sin\theta_1\cos^2\xi d\theta_1)^2\\[2mm]
+&\frac{1}{4}\cos^2\xi \sin^2\theta_1 \, G (d\phi_1+\frac{\gamma}{4}c(\rho)\cos\theta_2\sin 2\xi d\xi-\frac{\gamma}{4}c(\rho)\sin\theta_2\sin^2\xi d\theta_2)^2\\
+&\frac{1}{4}\sin^2\xi \sin^2\theta_2 \, G (d\phi_2+\frac{\gamma}{4}c(\rho)\cos\theta_1\sin 2\xi d\xi+\frac{\gamma}{4}c(\rho)\sin\theta_1\cos^2\xi d\theta_1)^2,\\
&\qquad \qquad G=(1+\gamma^2 \frac{\pi^2 \Delta}{h''^2}M)^{-1}, \quad M=s_{\theta_1}^2c_{\theta_2}^2c_{\xi}^2+s_{\theta_1}^2s_{\theta_2}^2+c_{\theta_1}^2s_{\theta_2}^2s_{\xi}^2. 
\end{split}
\end{equation}
The RR fields of the solution read
\begin{align}
	\label{eq:fluxesN=2}
	\tilde{F}_0&=-\frac{h'''}{2\pi},\nonumber\\
	\tilde{F}_2&=\frac{2((h'')^2-h'h''')}{h''}J+\frac{\gamma}{2}\left(\frac{\pi \Delta h'''}{(h'')^2}W+\left(\frac{\pi h'''(h')^2}{(h'')^2}+\frac{c(\rho)}{2h''}((h'')^2-h'h''')\right)Z_1\right)\nonumber\\
	\tilde{F}_4&=\pi \, d\left(h'+\frac{hh'h''}{\Delta}\right)\wedge\text{vol}(\text{AdS}_3)-\frac{4\pi h'''(h')^2}{(h'')^2}J\wedge J,\nonumber\\
	&-\frac{\gamma\pi \Delta }{2(h'')^3 M}\left[\left(3(h'')^2+\frac{h'^2h'''}{\Delta}(h'-2(\rho-k)h'')\right)Z_1\wedge Z_2\right.\nonumber\\
	&\left.+\left(((h'')^2-h'h''')Z_2+\gamma\pi^2M\frac{(h')^2h'''}{h''}Z_1\right)\wedge W\right].
\end{align}
where we have defined
\begin{align}
	Z_1&=c_\xi s_\xi\left(c_\xi s_\xi s_{\theta_1}s_{\theta_2}d\theta_1\wedge d\theta_2-2(s_{\theta_1}c_{\theta_2}c_{\xi}^2d\theta_1+c_{\theta_1}s_{\theta_2}s_{\xi}^2d\theta_2)\wedge d\xi\right),\nonumber\\
			Z_2&=\pi c_\xi s_\xi s_{\theta_1} s_{\theta_2}(c_{\theta_1}s_{\theta_2}s_\xi c_\xi d\theta_1-s_{\theta_1}(c_{\theta_2}c_\xi s_\xi d\theta_2+2s_{\theta_2}d\xi))\wedge d\psi , \nonumber\\
			W&=G c_\xi^2s_\xi^2\left(M d\phi_1\wedge d\phi_2-(c_{\theta_2}s_{\theta_1}^2c_\xi^2d\phi_1+c_{\theta_1}s_{\theta_2}^2s_\xi^2d\phi_2)\wedge d\psi+\gamma \, c(\rho)\left(M\;J-\frac{Z_2}{4\pi}\right)\right.\nonumber\\
			&\left.+\gamma^2\frac{c(\rho)^2}{16}M Z_1\right).
\end{align}
We have checked using Mathematica that this background is solution to massive IIA supergravity when $h''''=0$, in the absence of sources. 
From the solution it is clear that the isometry group of $\mathbb{CP}^3$ has been broken down to its subgroup U(1)$^3$, where one of the U(1)s -associated to $\psi$- 
is R-symmetry whilst the other correspond to global symmetries. In the massless limit (\ref{massless}) we recover an associated AdS$_4$ solution that was first constructed in \cite{Imeroni:2008cr}. 
It was shown in that reference, by using probe branes, that the moduli space of the deformed theory for rational $\gamma$, is an orbifold of the undeformed one. It will be interesting
to study this feature for the deformed solutions of this section, which are generalisation by D8 branes of those considered in the aforementioned reference. 

Before concluding this subsection, it is worth noticing that more solutions can be constructed by involving the external space in the deformation procedure. 
Perhaps the simplest case to consider is to write AdS$_3$ as a Hopf fiber of AdS$_2$ and take the fiber coordinate as one of the U(1)s in the procedure.
According to what we discussed in Section \ref{Gstr2}, if we consider in addition the internal coordiante $\psi$ as the second U(1), the solution then will break supersymmetry down to $\mathcal{N}=4$.
 On the other hand if we chose one of the  $\phi_i$, then the preserved supersymmetry will be $\mathcal{N}=2$. 
 For a single T-duality producing AdS$_2\times\mathbb{CP}^3$ see \cite{Conti:2023rul,Conti:2025djz}.


\section{The deformed bispinors and G-structures}\label{sec:defBi}

From the analysis of  section (\ref{Gstr1}), we showed the G-structure of the solutions defined on M$_5$ are uncharged under SU(2)$_R$ which 
ensured that by deforming the internal space M$_5$, the bispinors $\Psi^{(n)}_{\pm}$, $n=1,\ldots 4$ are all preserved. However, such an analysis 
does not tell us how the structure of the solution has transformed after the deformation. 
In this section we will study how the structure defined by the seed solutions is affected by the action of the TsT. 
We will see that depending on the choice of directions inside M$_5$, the structure will be type changing or slightly generalised by the deformation. In the same vein, for the solutions constructed in section (\ref{sol20}), we will
prove that the structure defined on the internal space M$_7$ is preserved. 

In order to achieve this, we will follow the results in Appendix \ref{sec:AppexA} on which it is explained the bispinors of the solution after a TsT transformation can be obtained by the right action of the operator\footnote{To avoid notational clutter we use $\Omega$ to denote this spinor rotation, which should not be confused with the complex three-form defining the SU(3) structure of the solutions}
$\Omega$ in equation (\ref{rotationp}) on the bispinors of the undeformed solutions. We would like to understand how this transformation acts on the structure 
defined by the undeformed solutions and we will do this in a case by case basis. 
In doing so, one can check that the bispinors so obtained solve the supersymmetric system (\ref{eq:BPSequations}) with associated supergravity fields. This is the case since for 
 isometries that commute with supersymmetry the twisted derivative $d_H$ commutes with the $\Omega$ action (see \cite{Butti:2007aq} for the untwisted case).

\subsection{The solution with dynamical SU(2)-structure on M$_5$} 

We will start by analising the bilinears of the solutions constructed in section \ref{sec:solIdentity}, namely, those in massive IIA. 
As we saw in that section, the transformation is performed along the two directions $(z_9,z_8)$ inside T$^4$ and as such 
the spinor transformation acting on the five-dimensional bilinears (\ref{eq:bilinearsSU(2)}) is $\Omega=\sqrt{G}(1+b\;\Gamma^{9 8})$,
where we have defined $b=\gamma \sqrt{g}$ and the functions $g, G$ are given in equation (\ref{back1def}). 
When we apply this spinor rotation on these bilinears we obtain 
\begin{align}\label{eq:bilinearsSU(2)dyn}
	\widetilde{\psi}_+^1&=\sqrt{G}(\psi_+^1+ib\;\psi_+^2),\qquad\qquad \widetilde{\psi}_+^2=\sqrt{G}(\psi_+^2+ib\;\psi_+^1),\nonumber\\
	\widetilde{\psi}_-^1&=\sqrt{G}(\psi_-^1+ib\;\psi_-^2),\qquad\qquad \widetilde{\psi}_-^2=\sqrt{G}(\psi_-^2+ib\;\psi_-^1).
\end{align}
where we see that the non-trivial part of the transformation acts as a type-changing bispinors since $\psi_{\pm}^{1,2}\leftrightarrow\psi_{\pm}^{2,1}$, which 
can be seen as a consequence of mirror symmetry.
These expression can be further simplified to give
\begin{align}\label{eq:bilinearsSU(2)dyn2}
	\widetilde{\psi}_+^1&=\sqrt{G}\;\psi_+^1\wedge e^{4ib \psi_+^2},\qquad\qquad \widetilde{\psi}_+^2=ib\sqrt{G}\; \psi_+^1\wedge e^{(-4ib^{-1} \psi_+^2)},\nonumber\\
	\widetilde{\psi}_-^1&=\sqrt{G}\;\psi_-^1\wedge e^{4ib \psi_+^2},\qquad\qquad \widetilde{\psi}_-^2=ib\sqrt{G}\;\psi_-^1\wedge e^{(-4ib^{-1} \psi_+^2)}.
\end{align}
Since the bispinors in equation (\ref{eq:bilinearsSU(2)dyn2}) are just the undeformed ones up to a point-dependent phase we conclude that the
deformed bispinors support a dynamical SU(2)-structure. 
Notice that a similar analysis was performed in \cite{Minasian:2006hv}. 
Had we used a different torus inside T$^4$ for the deformation -for instance
T$^2:(z_9,z_{10})$- the structure would have remained unchanged, namely dynamical SU(2),  but with bispinors mapping to themselves under the Clifford action by $\Omega$. This is the case since 
we are choosing directions which are paired on the structure forms (\ref{su2forms}).

\subsection{The solution with identity structure on M$_5$}
We now present the analysis of the bilinears for the massless IIA solution constructed in section \ref{masslessII}. In this case, we imposed that
the $\rho$ direction corresponds to an isometry of the solution and we used it together with an extra U(1)$\subset $T$^4$ to apply the TsT transformation.
Contrary to the case analysed in the previous subsection, we expect a type changing structure since the $\rho$ direction is now involved in the action of transformation 
$\Omega$ the action of which will be non-trivial in the structure forms. To be more precise by acting with $\Omega=\sqrt{G}(1+b\;\Gamma^{z_{10}\rho})$, and after massaging
the expression a bit, we obtain the following expressions 
\begin{align}
	\label{eq:bispinorsIdentity}
	\widetilde{\psi}_+^1&=\frac{\sqrt{G}}{4}e^{-ij-b\;w\wedge v},\qquad\qquad\qquad		\widetilde{\psi}_+^2=-\frac{\sqrt{G}}{4}u\wedge e^{-i j}\wedge(w-b\; v),\nonumber\\
		\widetilde{\psi}_-^1&=\frac{\sqrt{G}}{4}(v+b\;w)\wedge e^{-ij-b\;w\wedge v},\qquad
		\widetilde{\psi}_-^2=-\frac{\sqrt{G}}{4}u\wedge(w\wedge v+b\;e^{-i j}),
\end{align}
where $b=\gamma \sqrt{g}$ and $g, G$ are those in equation (\ref{seconddef}). The bilinears quoted in equation (\ref{eq:bispinorsIdentity}) are precisely the bilinears 
associated to an  identity structure on M$_5$, where the identity structure is spanned by the one forms $v$, $w$ and $u\equiv \frac{1}{2}\iota_{\bar{w}}\omega$, 
\begin{equation}\label{idstructureforms}
v=e^{\rho}, \qquad w=e^{10}-i e^9, \quad u=e^8-i e^7, 
\end{equation}
where the two-forms $j,\omega$ are given in terms of them as follows 
\begin{equation}
	j=\frac{i}{2}(w\wedge \bar{w}+u\wedge \bar{u}), \qquad \omega=w\wedge u.
\end{equation}
One can then easily check that the one-forms in (\ref{idstructureforms}) satisfy $v\cdot w=v\cdot u=0$, the real and imaginary parts of which define a vielbein for M$_5$. 

\subsection{A solution with SU(3)-structure and $\mathcal{N}=(2,0)$ supersymmetry}\label{sec:GSN=2}
In this section, we will study the structure supported by the solutions constructed in section (\ref{sol20}). We follow the same strategy as 
we did in the previous cases, namely study the Clifford action $\Omega$ on the bilinears parametrising the structure. At this point it is worth 
noticing that such an action will be only along the bispinors preserved after deformation, namely $\Psi^{(i)}_{\pm}, i=5,6$. In doing so, since
we are using the canonical frames in equation (\ref{framescp3}) we must be careful to bring the frames to the canonical form for T-duality for which the
direction along we will apply T-duality, is only in one frame component. We thus need to apply a couple of SO(6) rotations which act trivially along the 
$(e^{1},e^{2},e^4)$ and non-trivially  on the subspace spanned by
$(e^{3},e^{5},e^6)$ through SO(3) rotations, such that $\phi_1$ and $\phi_2$ are each in only one frame component\footnote{Actually, for T-duality along $\phi_1$ we only need one rotation to bring it into only one frame component, but by also achieving this for $\phi_2$ 
simplifies the analysis.}.
Explicitly, for $O, O'\in$ SO(3) rotations

\begin{equation}
\begin{array}{c}
O =
\begin{pmatrix}
\cos\kappa_1 & 0 & \sin\kappa_1 \\
0 & 1 & 0 \\
-\sin\kappa_1 & 0 & \cos\kappa_1
\end{pmatrix}, \\
\kappa_1=\arctan ( \sin\xi \cot \theta_1)
\end{array}
\quad
\begin{array}{c}
O' =
\begin{pmatrix}
1 & 0 & 0 \\
0& \cos\kappa_2 & -\sin\kappa_2 \\
0 & \sin\kappa_2 & \cos\kappa_2
\end{pmatrix},  \\
\kappa_2=\arctan (\frac{\sin\theta_1 \cos \xi \cot\theta_2}{\sqrt{\sin\theta^2_1+\cos\theta^2_1\sin^2\xi}})
\end{array}
\end{equation}
which gives the vielbeins 
\begin{equation}
\begin{split}\label{rotatede}
\hat{e}^3=&\frac{c_{\xi}}{\sqrt{s^2_{\theta_1}+c^2_{\theta_1} s^2_{\xi}}}\left(\frac{1}{2}d\phi_1 (s^2_{\theta_1}+c^2_{\theta_1} s^2_{\xi})-c_{\theta_1} s^2_{\xi}(d\psi-\eta_2)\right), \\
\hat{e}^5=&\frac{s_{\xi}}{4 \, c_{\xi} \sqrt{M(s^2_{\theta_1}+c^2_{\theta_1} s^2_{\xi})}}\left(c_{\xi} s_{\theta_1}M d\phi_2-s^3_{\theta_1}c^2_{\theta_2}c^3_{\xi}d\psi \right), \\
\hat{e}^6=& \frac{s_{2\xi} s_{\theta_1}s_{\theta_2}}{8 \sqrt{M}} d\psi, 
\end{split}
\end{equation}
where $M$ was defined in equation (\ref{definitionsH}). Each of the rotations induces a corresponding transformation on the bispinor that can be extracted from the relation $\Omega^{-1}\Gamma^a \Omega=O^a{}_b\Gamma^b$.
We find
\begin{equation}
\Omega_{O}=\cos\frac{\kappa_1}{2}+\sin\frac{\kappa_1}{2}\Gamma^{3 6}, \quad  \Omega_{O'}=\cos\frac{\kappa_2}{2}-\sin\frac{\kappa_2}{2}\Gamma^{56},
\end{equation}
where the $\Gamma$-matrices are all flat. These matrices have a left-right action on the polyforms,
\begin{equation}
\Psi_{\pm}^{(I)}\rightarrow\Omega_{O'}\Omega_{O}\Psi_{\pm}^{(I)}(\Omega_{O'}\Omega_{O})^{-1}, \quad I=5,6,
\end{equation}
from which we obtain the transformed bispinors by the action of $\Omega=\sqrt{G}(1+b \hat{\Gamma}^{35})$, where the hat in the gamma matrices indicates they are
flat with respect to the frames (\ref{rotatede}), $G$ is given in equation (\ref{definitionsH}) and $b=\gamma \pi \sqrt{\Delta\, M}/h''$.
After a considerable number of simplifications,
we find the bispinors
 \begin{align}\label{bilinearsm}
	\tilde{\Psi}_{+}^{(5)}&=\Psi_{+}^{(5)}+\frac{bc_\alpha}{8}\;\left[s_{\beta_2}-c_{\beta_2} \mathcal{J}_b-\frac{c_\alpha^2 c_{\beta_2}^2}{\varkappa}\left(\frac{s_{\beta_2}}{2} -\frac{c_{\beta_2}}{6}\mathcal{J}_b\right)\wedge\mathcal{J}_b\wedge\mathcal{J}_b-\frac{c_\alpha^2 c_{\beta_2}^3}{\varkappa^{3/2}}\text{Re}[\Omega_b]\wedge V_b\right],\nonumber\\
	\tilde{\Psi}_{-}^{(5)}&=\Psi_{-}^{(5)}-\frac{bc_\alpha}{8}\;\left[\left(c_{\beta_2} + s_{\beta_2}\mathcal{J}_b-\frac{c_\alpha^2 c_{\beta_2}^2}{\varkappa}\left(\frac{ c_{\beta_2}}{2} +\frac{s_{\beta_2}c_\alpha^2 c_{\beta_2}^2}{6\varkappa}\mathcal{J}_b\right) \wedge\mathcal{J}_b\wedge\mathcal{J}_b\right)\wedge V_b -\frac{ c_{\beta_2}}{\sqrt{\varkappa}}\text{Im}[\Omega_b]\right],\nonumber\\
	&\textrm{where}\qquad c_\alpha=\frac{s_{2\xi}c_{\psi}s_{\theta_1}s_{\theta_2}}{\sqrt{M}}, \qquad \varkappa=1-c_\alpha^2s_{\beta_2}^2,
\end{align}
where we have introduced a new set of forms $(\mathcal{J}_b,\Omega_b, V_b)$,  parametrising an SU(3)-structure (see below) for the deformed solution
satisfying \eqref{eq:SU(3)restrictionsonM7}. Notice we have only quoted the bispinor associated to $I=5$. Similar expressions in terms of an SU(3)-structure
 can be obtained for $\tilde{\Psi}_{\pm}^{(6)}$. 

Moreover, following the results in \cite{Butti:2007aq}, the bilinears in equation (\ref{bilinearsm}) can also  be written in terms of SU(2)-structure forms as, 
\begin{align}
	\tilde{\Psi}_{+}&=\frac{1}{8}\;\text{Re}\left[e^{i\beta_2}e^{-i(j+U_1\wedge U_2)}\wedge (1-i bc_\alpha\; e^{-\tan\alpha\; \omega})-(e^{i\beta_1}U\wedge\omega-i bs_\alpha\; e^{-i\beta_1}\bar{U}\wedge e^{i j+\cot\alpha\;\bar{\omega}})\wedge V\right],\nonumber\\
	\tilde{\Psi}_{-}&=\frac{1}{8}\;\text{Im}\left[e^{i\beta_2}e^{-i(j+U_1\wedge U_2)}\wedge (1-i bc_\alpha\; e^{-\tan\alpha\; \omega})\wedge V+e^{i\beta_1}U\wedge\omega+ibs_\alpha\; e^{-i\beta_1}\bar{U}\wedge e^{ij+\cot\alpha\; \bar{\omega}}\right],
\end{align}
where $U$ and $\omega$ are complex one- and two-forms respectively, whereas $j$ is a real two-form. These forms describe an SU(2)-structure \eqref{eq:SU(2)onM4} in 4-dimensions, and their relation with the SU(3) invariant quantities is given by (see also \cite{Macpherson:2021lbr}), 
\begin{align}
	&\mathcal{J}=j+U_1\wedge U_2,\qquad \qquad \Omega=U\wedge\omega,\qquad\qquad \omega=\omega_R+i\omega_I,\nonumber\\
	&U=U_1+i U_2,\qquad\qquad \iota_{U_1}U_2=0,\qquad\qquad  \iota_{U_2}U_2=\iota_{U_1}U_1=1,
\end{align}
while their relation with the SU(3) deformed forms is the following,
\begin{align}
	V_b=&V-\frac{\tan\alpha}{c_{\beta_2}}\textrm{Re}[e^{i\beta_1}U],\nonumber\\
	\Omega_b=&\frac{\sqrt{\varkappa}}{ c_{\beta_2}}\left[\frac{\tan\alpha\; (s_\alpha^2+c_\alpha^2c_{\beta_2}^2)}{\varkappa c_{\beta_2}}\textrm{Im}[e^{i\beta_1}U]\wedge(s_{\beta_2}j+ s_{\beta_2}\tan\alpha\;\omega_I-\frac{c_{\beta_2}}{c_\alpha s_\alpha}\omega_R)\right.\nonumber\\
	&\left.+(\sec\beta_2\;\tan\alpha\; V+\textrm{Re}[e^{i\beta_1}U])\wedge (\tan\alpha\; j-\omega_I)+\frac{i}{c_\alpha^2c^2_{\beta_2}}\left(s_\alpha V\wedge(c_\alpha c_{\beta_2}\omega_R+s_\alpha s_{\beta_2}U_1\wedge U_2)\right.\right.\nonumber\\
	&\left.\left.+c_\alpha c_{\beta_2}^2\textrm{Im}[e^{i\beta_1}U]\wedge(s_\alpha\;j-c_\alpha\omega_I)-\textrm{Re}[e^{i\beta_1}U]\wedge\left((s_\alpha^2-c_{\beta_2}^2)\omega_R+s_\alpha c_{\beta_2}s_{\beta_2}(c_\alpha\;j+s_\alpha\omega_I)\right)\right)	\right], \nonumber\\
	\mathcal{J}_b=&\mathcal{J}+\tan\alpha(\omega_I+\tan\beta_2\omega_R-\sec\beta_2\; \textrm{Im}[e^{i\beta_1}U]\wedge V).
\end{align}
The explicit  expressions for $j$, $U$ and $\omega$ are somewhat unwieldy, we quote the explicit expression in equation \eqref{eq:jUw}.


\subsection{Supersymmetric branes after deformation}\label{susybranesdef}
In this section we will study D-brane embeddings in the deformed solution of section \ref{sec:solIdentity} following the analysis of section \ref{susybranes} for the undeformed solution.
As we discussed in section \ref{sec:solIdentity}, the quantisation of the fluxes for the background of equations (\ref{back1def})-(\ref{defRR1}) tell us that on top of the D-branes intersecting 
in the way described in Table \ref{branesetup}, new D-branes appear  as a consequence of the deformation if the deformation parameter $\gamma$ is
rational. More concretely, we have the following mapping
\begin{equation}\label{mapping}
(\text{N}_{\text{D}8},\text{N}_{\text{D}6}, \text{N}_{\text{D}4},\text{N}_{\text{D}2}) \xrightarrow{\text{TsT}} (\text{N}_{\text{D}8},\text{N}_{\text{D}6}+\gamma \text{N}_{\text{D}4}, \text{N}_{\text{D}4}+\gamma \text{N}_{\text{D}2},\text{N}_{\text{D}2} )
\end{equation}
where the new $\tilde{\text{D}}4$ and $\tilde{\text{D}}6$ branes span the  submanifolds $ (\text{Mink}_2,\rho, \text{T}^2)$ and $(\text{AdS}_3,\text{S}^2, \text{T}^2)$ respectively.
From the worldsheet perspective, this can be understood since the D6 and D8 branes have Neumann (NN) boundary conditions along the two torus whilst D4 and D2 Dirichlet ones (DD). 
The deformation is such that it maps NN$\rightarrow$NN and DD$\rightarrow$NN \cite{Imeroni:2008cr},  in the latter the corresponding D-brane wrapping T$^2$.
Let us study their embeddings separately. 

\begin{itemize}
\item A D2 brane along $\Sigma^{(3)}=(\text{Mink}_2$, $\rho$)

The DBI and WZ Lagrangian densities for the D2 brane using the C-potentials in equation (\ref{cpotdef}) are 
\begin{equation}
e^{-\tilde{\Phi}}\sqrt{\mid \tilde{g}+\tilde{B}_2\mid} \Big\vert_{\Sigma^{(3)}}=\frac{1}{2}r^2 \sqrt{\frac{h_8 g_1}{h_4 G} }, \quad \tilde{C}_3-\tilde{C}_1\wedge \tilde{B}_2\Big\vert _{\Sigma^{(3)}}=\frac{1}{2}r^2 \sqrt{\frac{h_8}{h_4} g_1} ,
\end{equation}
where we observe an extra $1/\sqrt{G}$ factor in the DBI term due to the deformed dilation, which generates a gravitational potential that is not cancelled by the WZ term. Therefore 
the D2 brane feels an attractive force that brings it to the Poincare horizon at $r=0$. This is contrary to what happens to the D2 brane in the undeformed background since it
was BPS without imposing any condition. Notice we can also be tempted to move the D2 branes to the point where the two-torus shrinks to zero size such that $G=1$, where G is given in equation (\ref{back1def}). 
We can immediately convince ourselves that this is not possible since the D2 wraps the $\rho$-interval. 

\item A D6 brane along $\Sigma^{(7)}=(\text{Mink}_2$, $\rho$, T$^4$)

For this D brane an easy computation shows that 
\begin{equation}
e^{-\tilde{\Phi}}\sqrt{\mid \tilde{g}+\tilde{B}_2]\mid }\Big\vert_{\Sigma^{(7)}}=\frac{1}{2}r^2 \sqrt{\frac{h_4}{h_8} g_1},\quad \tilde{C}_7-\tilde{C}_5\wedge \tilde{B}_2 \Big\vert_{\Sigma^{(7)}}=-\frac{1}{2}r^2 \sqrt{\frac{h_4}{h_8} g_1},
\end{equation}
from which we see the D6 brane is still BPS after the deformation. 
From this analysis, we learn that if a D-brane initially wraps the two-torus along which we performed the deformation and is BPS,
it will remain BPS after the deformation. On the other hand if the two-torus is not part of the world-volume of the brane then the no-force condition will be violated
unless we can move the corresponding brane to a point where the T$^2$ shrinks to zero size.  

\item A $\tilde{\text{D}}4$ brane along (Mink$_2$, $\rho$, T$^2$)

Notice this is one of the D-branes that appeared as a consequence of the deformation. 
The DBI and WZ terms for this brane are given by 
\begin{equation}
e^{-\tilde{\Phi}}\sqrt{\mid \tilde{g}+\tilde{B}_2\mid} \Big\vert_{\Sigma^{(3)}\times \text{T}^2}=\frac{1}{2}r^2 \sqrt{\frac{h_8 g_1}{h_4 G} }, \quad \tilde{C}_5-\tilde{C}_3\wedge \tilde{B}_2+\frac{1}{2}\tilde{C}_1\wedge \tilde{B}_2\wedge \tilde{B}_2\Big\vert _{\Sigma^{(3)}\times \text{T}^2}=0 ,
\end{equation}
so there is no charge associated to the D6 brane, which could have been inferred from the charge distribution quoted in (\ref{mapping}). In order to show that this brane is 
stable we ought to switch on a magnetic world-volume flux along the 
T$^2$ wrapped by the brane of magnitude $f_{z_9\, z_8}=1/\gamma$. In doing so we obtain 
\begin{equation}
e^{-\tilde{\Phi}}\sqrt{\mid \tilde{g}+\tilde{B}_2\mid} \Big\vert_{\Sigma^{(3)}\times \text{T}^2}=\frac{1}{\gamma}\frac{r^2}{2} \sqrt{\frac{h_8}{h_4} g_1}, \quad -\tilde{C}_3\wedge f_2\Big\vert_{\Sigma^{(3)}\times \text{T}^2} =\frac{1}{\gamma}\frac{r^2}{2} \sqrt{\frac{h_8}{h_4} g_1},
\end{equation}
showing that the brane is stable due to induced charge of a D2 brane.

Since the remaining branes turn out to be flavour branes we will analyse their stability using the calibrations in the same way we did in section (\ref{susybranes}) for the undeformed solutions. Let us start with the 
D4 and D8 branes which are part of the undeformed solution. 

\item A D4 brane along $(\text{AdS}_3, S^2)$ 

We recall that this D brane corresponds to a supersymmetric embedding for $\gamma=0$. 
In the deformed solution we impose the calibration condition of equation  (\ref{cal}) with fields and bispinor replaced by the deformed ones.
Explicitly, the calibration form and DBI Lagrangian
density for the corresponding brane placed at $\rho=2\pi k$ turn out to be
\begin{equation}
\begin{split}
e^{-\Phi}\sqrt{\mid \tilde{g}+\tilde{B}_2\mid }\Big\vert_{\Sigma^{(5)}}=& \frac{1}{\sqrt{G}}\frac{u^2}{4 h_4}r \sin\theta,\\
\tilde{\Psi}^{(\text{cal})}_{D4}=&\frac{u^2 }{4 h_4}\, \text{vol}(\text{AdS}_3)\wedge \text{vol} (\text{S}^2),
\end{split}
\end{equation}
where, as we saw before, since the brane is not wrapping T$^2$ an extra contribution is generated from the DBI term as consequence of the  deformation. 
Therefore, the D4 brane does not correspond to a supersymmetric embedding. 

\item A D8 brane along $(\text{AdS}_3, S^2, \text{T}^4)$

Following the same reasoning, since this brane is wrapping T$^2$ we anticipate it will correspond to a supersymmetric embedding in the deformed background.
Indeed a quick computation shows that for a D8 brane sitting at $\rho=2\pi k$ 
\begin{equation}
\begin{split}
e^{-\tilde{\Phi}}\sqrt{\mid \tilde{g}+\tilde{B}_2\mid }\Big\vert_{\Sigma^{(9)}}=&\frac{u^2}{4 h_8}r \sin\theta, \\
\tilde{\Psi}^{(\text{cal})}_{D8}=&\frac{u^2 }{4 h_8}\, \text{vol}(\text{AdS}_3)\wedge \text{vol} (\text{S}^2),
\end{split}
\end{equation}
showing that the embedding of the D8 brane is supersymmetric.

\item A $\tilde{\text{D}}6$ brane along $(\text{AdS}_3, S^2, \text{T}^2)$
This case is similar to the $\tilde{\text{D}}4$ brane embedding studied above. Namely, from the charge distribution (\ref{mapping}) we anticipate this brane will have induced D4 charge.  
A straightforward computation shows that if $f_2=0$ the calibration condition is not satisfied. We then switch on a magnetic world-volume flux $f_{z_9, z_8}=1/\gamma$ to show that  
\begin{equation}
\begin{split}
e^{-\tilde{\Phi}}\sqrt{\mid \tilde{g}+\tilde{B}_2\mid }\Big\vert_{\Sigma^{(5)}\times \text{T}^2}= \frac{1}{\gamma}\frac{u^2}{4 h_4}r \sin\theta=
\tilde{\Psi}^{(\text{cal})}_{\tilde{\text{D}}6}=\frac{1}{\gamma}\Psi^{(\text{cal})}_{D4},
\end{split}
\end{equation}
such that the $\tilde{\text{D}}6$ are calibrated.  
\end{itemize}

From the analysis in this subsection, we conclude that not all branes inferred from charge quantisation correspond to BPS configurations. More importantly, since the D4 branes
are not supersymmetric embeddings, we cannot use them as defect branes to construct global solutions with a bounded interval. However, we have seen that the
$\tilde{\text{D}}6$ -which wrap T$^2$- can be used as defects branes instead. This should not come as a surprise since we have seen the $\tilde{\text{D}}4$ branes correspond to the D2 that polarised
due to a Myers-like effect. The same logic follows for the $\tilde{\text{D}}6$ branes. 
One may then consider the brane configuration $\tilde{\text{D}}4$-$\tilde{\text{D}}6$-D6-D8-NS5 to analyse the effect of the deformation on the field theory, following what was done for the undeformed 
solution in \cite{Lozano:2019zvg}. Compared to this, we have seen that the charges of D2 and D4 are modified by a factor $\gamma$ (see equation \ref{massless}) which ought to be rational. 
For the choice $\gamma=1/n$, we then start with $n\alpha_k$ D2 branes and $n\beta_k$ D4 branes, which is reminiscent of an orbifold action,  similar to the solution studied in \cite{Faedo:2020lyw}.
Since the charges of colour and flavour branes are in one to one correspondence with the rank of gauge and flavour groups of the quiver field theory respectively, 
we have for gauge and flavour groups that SU$(n\alpha_k)$ and SU$(n\beta_k)$ turn into SU$(n \alpha_k) \rightarrow \Pi_{i=1}^n \text{SU}(\alpha_k)^{(i)}$ and SU$(n \beta_k) \rightarrow \Pi_{i=1}^n \text{SU}(\beta_k)^{(i)}$
respectively. One may wonder under which conditions the field theory is still well-defined after this operation. A hint towards answering this question can be obtained by analysing the cancellation of anomalies for the 
dual quiver field theory, which for the $\mathcal{N}=(4,0)$ theories of this class implies that
for the SU$(\beta_k)$ and  SU$(\alpha_k)$ gauge groups the following constraint among the ranks of the groups must be satisfied \cite{Lozano:2019zvg}
\begin{equation}
2Q^{(k)}_{\tilde{\text{D}}4}=\Delta Q^{(k)}_{\tilde{\text{D}}6}, \quad\qquad 2Q^{(k)}_{\text{D}6}=\Delta Q^{(k)}_{\text{D}8},
\end{equation}
where $\tilde{\text{D}}6$ and $\text{D}8$ is the number of  $\tilde{\text{D}}6$ and $\text{D}8$ branes in the $k$-th interval respectively.  Since the orbifolding involves redefining 
the function $h_4$ in Table \ref{piecewiseh} as $h_4\rightarrow n\, h_4$, it is clear the cancellation of anomalies still holds. 
 Moreover, since the deformation we have considered is marginal one may be able to compute the 
central charge read off from the quiver of the deformed solution and find the same result to that of the undeformed theory.  We leave the details of this computation for
a future work. 
 With this, we conclude the technical part of our work.
To those who made it to these last lines -your patience (and endurance) is truly appreciated!

\section{Conclusions}\label{conclusions}
In this work we have constructed new solutions in massive and massless IIA supergravity which were obtained by applying a TsT transformation
along internal space  directions of the solutions, preserving different amount of supersymmetries. 
We first started by introducing the seed solutions in massive IIA supergravity and discussed some of their properties. In particular, the way sources can be added to the solutions
such that we can achieve a compact internal space. This fact enables us to study the quantisation of the fluxes for the solutions. Doing this for the family of solutions
 in section (\ref{solution1}), gives rise to a D2-D4-NS5-D6-D8 brane configuration described in Table (\ref{branesetup}). The same can be achieved for the other solutions, but we 
 focused only on this solutions as a representative example. 
 Subsequently, we discussed the supersymmetries and G-structures each solution support using bispinor techniques. In particular, we wrote down
 the bispinors in terms of the G-structure forms each of the solutions support. We also studied supersymmetric BPS D-brane embeddings
 in the conventional way using the action for the Dp branes and for the case of source branes, the formalism of calibrations in terms of the
 polyform bilinears. 
 We then proceed to construct the new solutions using the TsT transformation. For this, we discussed the possibilities of deforming the internal space of the solutions
 whilst preserving full or partial supersymmetry of the original solutions. Choices other than the ones we discussed will lead to non-supersymmetric solutions--unless one involves AdS$_3$.
 After that, we proceed to characterise the solutions in terms of the G-structures they preserve after the deformation is applied directly on the bispinors.
 After lengthy manipulations, we managed to write down these bispinors in terms of a new set of G-structure forms characterising the solution. 
 We showed explicitly that the G-structures preserved can be type changing or not depending on the directions one chooses to apply the deformation. 
 Finally, we analysed BPS brane embeddings for the deformed solutions of section (\ref{sec:solIdentity}) in order to compare the analysis of the undeformed solutions. 
 We verified some of the branes underlying the undeformed solution do not satisfy the no-force condition due to a gravitational potential generated by the deformation.
 Moreover, new Dp+2 branes which are polarised Dp branes are shown to be stable. This gives rise to a "new" brane configuration which correspond to a different-looking
 intersection associated to the dual deformed solution. We discussed how the deformation modifies the charges and as such the gauge groups/flavour groups of the
 dual field theory such that anomalies are trivially canceled. It would be interesting to study more field theory aspects of the solutions constructed here, but we leave that for
 a future work. 
 Thats all folks!

\acknowledgments
We would like to thank Niall Macpherson for very useful discussions. We also thank Yolanda Lozano and Alessandro Tomasiello for comments on the manuscript. 
 AR is supported by the Postdoctoral Fellowship Junior Marie Curie of the Research Foundation - Flanders (FWO) grant 12ACZ25N. 
 The work of SZ is supported by the grant “Dualitites and higher order derivatives” (GA23-06498S) of the Czech Science Foundation (GACR) .


\appendix
\section{TsT transformation and spinors}\label{sec:AppexA}

In this appendix we will present the transformation rules of the RR fluxes, associated C-potentials and the polyform bilinears under the TsT. 

We start by considering a supergravity background admitting two commuting U(1) isometries, generated by $k_{u}$ and  $k_{v}$,  acting by shifts along the
coordinates $u,v$,  respectively. We introduce the following frame adapted to these isometries  
\begin{equation}\label{natframe}
e^A=(e^a{}_{\mu} dx^{\mu}, e^{C_1}(du+A_1),e^{C_2}(dv+A_2)), \quad a=1,\ldots , 8, 
\end{equation}
where the warping factors and one-forms $A_i,\, i=1,2$ are independent of the $(u,v)$ coordinates such that the vectors $k_{u}$ and  $k_{v}$ are Killing.
This condition ensures consistency with T-duality and must hold for all supergravity fields of the solution. It is worth noting that we can be less restrictive by 
imposing that, in a certain gauge, the Kalb-Ramond two-form and the C-potentials satisfy $\mathcal{L}_{k_v}B=d\Lambda$, $\mathcal{L}_{k_v}C=d\overline{\Lambda}$ \cite{Grana:2008yw}.

The  vielbein components and dilaton after a T-duality along the direction $v$ read\footnote{as a slight abuse of notation we denote the T dual coordinate again $v$} \cite{Bergshoeff:1994cb}
\begin{equation}\label{frame}
\begin{split}
\hat{e}^{m}{}_{v}=\frac{1}{g_{vv}}e^{m}{}_{v},& \quad  \hat{e}^{m}{}_{\sigma}=e^{m}{}_{\sigma}-\frac{1}{g_{vv}}(g_{\sigma v}+B_{\sigma v})e^{m}{}_{v}, \quad \sigma\neq v, \\
&\quad \quad \quad \quad e^{\hat{\Phi}}=e^{\Phi-C_2}. 
\end{split}
\end{equation}
On the other hand, the transformation of the RR  sector can be ascertained by first noting that, under the Clifford map,  the RR polyform $F$ becomes a bispinor
\begin{equation}
\frac{1}{k!}F_{ a_1,\ldots ,a_k}e^{a_1}\wedge \ldots \wedge e^{a_k}\leftrightarrow \frac{1}{k!}F_{ a_1,\ldots ,a_k}\Gamma^{a_1\ldots a_k}\equiv \slashed{F}_k, 
\end{equation}
where $\Gamma^{a_1\ldots a_k}$ are antisymmetric products of flat space gamma matrices. 
In the frame (\ref{natframe}), 
the T-dual fluxes are obtained by Clifford multiplication
\begin{equation}\label{trule}
e^{\hat{\Phi}}\, \hat{\slashed{F}}=e^{\Phi}\, \slashed{F}\, \Gamma^v,  \quad \slashed{F}\Gamma^{a}=(-1)^{k}(e^a\wedge -\iota ^a)\slashed{F}.
\end{equation}
The second step involves shifting the coordinate $u$ as $u \rightarrow u+\gamma \, v$. In doing so,
the frames must be rotated back to their canonical form (\ref{frame}) via an SO(2) rotation,
\begin{equation}\label{roto}
 e^a=O^a{}_b e^b, \quad O=\frac{1}{\vert O\vert} \begin{pmatrix}
1 & -\gamma \sqrt{g} \\
\gamma \sqrt{g} & 1
\end{pmatrix},
 \end{equation}
where $g=e^{C_1+C_2}$ is the determinant of the two torus spanned by $(u,v)$.
 This rotation induces a transformation on the flux bispinor that can be obtained by solving $\Omega^{-1}\Gamma^a\Omega=O^a{}_b\Gamma^b$.  
 The transformation is
 \begin{equation}
 \hat{\slashed{F}}\rightarrow \Omega^{-1}\hat{\slashed{F}} \Omega . 
 \end{equation}
Successive application of the 
transformations (\ref{frame}) and (\ref{trule}) for a second T-duality along $v$ results in 
\begin{equation}\label{rotationf}
e^{\tilde{\Phi}}\, \tilde{\slashed{F}}=e^{\Phi}\, \slashed{F}\, \Omega, \quad \Omega=\sqrt{G}\left(1+\gamma \sqrt{g}\, \Gamma^{vu}\right),
\end{equation}
where $\tilde{\Phi}$ is the dilaton after TsT satisfying $e^{\tilde{\Phi}-\Phi}=\sqrt{G}$, with $G=(1+\gamma^2 g)^{-1}$.  It is important to highlight that, in order to
obtain the transformation (\ref{rotationf}) we had to undo the rotation (\ref{roto})  and thus a rotation  on the frames $\tilde{e}^a\rightarrow  (O^{\mathrm{T}})^a{}_b \tilde{e} ^b$, must be applied.
Moreover, 
in certain situations it is more convenient to know the expressions for the C-potentials associated to the transformed fluxes $\tilde{F}$.
Their transformation rule can be directly obtained from equation (\ref{rotationf}) and reads
\begin{equation}\label{rotationc}
e^{\tilde{\Phi}}\, \tilde{\slashed{C}}=e^{\Phi}\, \slashed{C}\, \Omega.
\end{equation}

Finally, and more importantly for the purposes of the present work, one can also ascertain the transformation of the polyform bilinears paralleling the discussion above,
 since they become bispinors via the Clifford map. The transformation turns out to be
\begin{equation}\label{rotationp}
\Psi\rightarrow \tilde{\Psi}=\Omega\, \Psi, \quad \Gamma^a \Psi=(e^a\wedge +\iota ^a)\Psi . 
\end{equation}
Therefore, if the $(u,v)$ are supersymmetry preserving isometries, $\mathcal{L}_{k_v}\Psi=\mathcal{L}_{k_u}\Psi=0$,  the TsT polyform bilinears $\tilde{\Psi}$
solve the BPS system (\ref{eq:BPSequations}) for the TsT supergravity fields. If, in addition, we solve the Bianchi identities for the fluxes without sources, 
it follows that any supersymmetric solution is mapped to another supersymmetric solution. 


\section{Page charges and Bianchi identities}\label{apppage}

In this appendix we consider the quantisation of the Page charges for the solutions discussed in Sections \ref{solution1}. The page charge
for a Dp-brane is given by
the expression 
\begin{equation}
Q_{Dp}=\frac{1}{(2\pi)^{7-p}} \int _{\Sigma_{8-p}} \hat{f}_{8-p}\,  d\vec{x},
\end{equation}
where $\Sigma$ is a compact cycle in the geometry and $\hat{f}$ is the magnetic component of the page flux $\hat{F}=Fe^{-\mathcal{F}}$
In order to proceed we will consider the piece-wise linear functions given in Table \ref{piecewiseh} parametrised in terms of constants $\alpha_k, \beta_k, \mu_k$ and $\nu_k$.

\begin{table}[h!]
\centering
 \begin{tabular}{c||c c c|ll|lc c c|c lc c c ll c c}
  & $0\leq \rho\leq 2\pi$ & $2\pi j\leq \rho\leq 2\pi(j+1)$ & $2\pi P\leq \rho\leq 2\pi(P+1)$  \\ [0.5ex] 
 \hline\hline 
 $h_8$ & $ \frac{\nu_0 }{2\pi}\rho$ & $ \mu_j+\frac{\nu_{j}}{2\pi}(\rho-2\pi j)$ & $ \mu_P - \frac{\mu_P}{2\pi}(\rho-2P\pi )$\\ 
 \hline
 $h_4$ & $ \frac{\beta_0 }{2\pi}\rho$ & $ \alpha_j+\frac{\beta_{j}}{2\pi}(\rho-2\pi j)$ & $ \alpha_P - \frac{\alpha_P}{2\pi}(\rho-2 P\pi )$  \\ 
 \hline 
  $u$ & &   $\frac{b_{0}}{2\pi}\rho$ & \\
  \hline
\end{tabular}
\caption{An example of piece-wise linear functions $h_4$ and $h_8$ where each sub-interval is of length $2\pi$.}
\label{piecewiseh}
\end{table}
Continuity of the fields across intervals imposes the following relations among the
various constants 
\begin{equation}
\alpha_{k}=\sum_{j=0}^{k-1}\beta_j,\quad \mu_{k}=\sum_{j=0}^{k-1}\nu_j. 
\end{equation}

The Page fluxes associated to the RR fluxes in equation (\ref{back1fluxes}) are
 \begin{equation}
       \begin{split}
       \hat{f}_0&= h'_8, \quad 
       \hat{f_2}=-\frac{1}{2}(h_8-h_8'(\rho-2\pi k))\,  \text{vol}(\text{S}^2),\\
         \hat{f}_4&=h'_4\text{vol}(\text{T}^4), \quad  \hat{f_6}=\frac{1}{2}(h_4-h_4'(\rho-2\pi k))\,   \text{vol}(\text{S}^2) \wedge \text{vol}(\text{T}^4). 
       \end{split}
       \end{equation}
    In the interval $\rho\in [k,k+1]$ we obtain the following charges of Dp and NS5 branes
\begin{equation}\label{page1}
\begin{split}
Q^{(k)}_{_{\text{NS}5}}=&\frac{1}{(2\pi)^2}\int_{ (\rho,S^2)}H_3=1,  \quad  Q^{(k)}_{D8}=2\pi h_8'=\nu_k, \quad Q^{(k)}_{D6}=\frac{1}{2\pi}\int_{ S^2}\hat{f}_2=\mu_k\\
&Q^{(k)}_{D4}=\frac{1}{(2\pi)^3}\int_{ T^4}\hat{f}_4=\beta_k,\quad Q^{(k)}_{D2}=\frac{1}{(2\pi)^5}\int_{(S^2, T^4)}\hat{f}_6=\alpha_k, 
\end{split}
\end{equation} 
where for them to be quantised we impose all constants to be integers. 
Since the corresponding branes wrap orthogonal cycles to the fluxes they are sourced by we conclude the brane configuration underlying the solution of Section (\ref{solution1}) 
corresponds to a  D2-D4-NS5-D6-D8 brane intersection in $\text{Mink}_2\times \mathbb{R}^3\times \text{T}^4\times I_{\rho}$ where D2 branes lie along $\text{Mink}_2\times I_{\rho}$, 
D6 in $\text{Mink}_2\times \text{T}^4\times I_{\rho}$,  D4 in $\text{Mink}_2\times\mathbb{R}^{3}$ smeared in  T$^4$, D8 in $\text{Mink}_2\times \mathbb{R}^3\times \text{T}^4$, the last two
having an enhancement to AdS$_3$ in the near horizon limit, and finally the NS5 are wrapping $\text{Mink}_2\times\text{T}^4$.  On the other hand, a direct computation of the Bianchi identities gives 
\begin{align}
        d\hat{f}_0=&2\pi h''_{8}d\rho, \qquad \qquad 
        d\hat{f_2}=0,\\
        d\hat{f_4}=&h''_{4}d\rho\wedge \text{vol}(\text{T}^4), \quad 
        d\hat{f}_6=0.
        \end{align}
        Taking into account the functions $h_4$ and $h_8$ from Table \ref{piecewiseh} we see that their second derivatives vanish everywhere except at special points where there is a change in slope
         and $h''_{4,8}\sim \delta(\rho-\rho_0)$ giving rise to a source term for the corresponding brane. From this analysis we conclude 
         that D4 and D8 branes correspond to flavour branes whilst D2 and D6 to colour branes.

When the deformation is turned on, new flux components arise sourcing additional branes provided their charges are quantised. Using the RR fluxes in equation (\ref{defRR1}) we find
the following Page fluxes
 \begin{equation}\label{deformedf}
        \begin{split}
        \hat{\tilde{f}}_2=&-\frac{1}{2}(h_8-h_8'(\rho-2\pi k))\text{vol}(S^2)+\gamma\, h_4' dz_7\wedge dz_{10},\\
        \hat{\tilde{f}}_4=&h'_4\text{vol}(\text{T}^4)
        -\frac{\gamma}{2}(h_4-h_4'(\rho-2\pi k))\text{vol}(S^2)\wedge dz_7\wedge dz_{10},\\
        \hat{\tilde{f}}_6=&-\frac{1}{2}(h_4-h_4'(\rho-2\pi k))\text{vol}(S^2)\wedge \text{vol}(\text{T}^4).
        \end{split}
        \end{equation}
Therefore, aside from the branes already quoted in (\ref{page1}), the following brane charges arise
\begin{equation}\label{page2}
Q^{(k)}_{\tilde{\text{D}}6}=\frac{1}{2\pi}\int_{ z_7\, z_{10}}\hat{\tilde{f}}_2=\gamma \, Q^{(k)}_{D4},\quad Q^{(k)}_{\tilde{\text{D}}4}=\frac{1}{(2\pi)^3}\int_{(S^2,z_7\, z_{10})}\hat{\tilde{f}}_4=\gamma \, Q^{(k)}_{D2}
\end{equation}
which are quantised provided $\gamma Q^{(k)}_{\text{D}4}, \gamma Q^{(k)}_{\text{D}2}\in \mathbb{Z}$ implying $\gamma$ must be rational. Moreover, the Page fluxes under which these branes are charged satisfy 
\begin{align}\label{bideformed}
        d\hat{\tilde{f}}_2=&\gamma h''_4 d\rho\wedge dz_7\wedge dz_{10}, \quad 
        d\hat{\tilde{f}}_4=0.
        \end{align}
From the above we immediately see that the $\tilde{\text{D}}6$ and $\tilde{\text{D}}4$ branes generated by the deformation correspond to flavour and colour branes respectively. 

Moreover, as we pointed out in Section (\ref{solution1}) the D4 and D8 sources in the seed solution were used to construct an infinity family of solutions with a compact internal space. This was achieved by placing 
these branes at special points along the $\rho$ interval allowing the linear functions $h_4, h_8$ to be globally piece-wise convex linear functions. In doing this the sources ought to solve the modified 
Bianchi identities as well as be supersymmetric embeddings \cite{Lozano:2019emq}. For the deformed solutions, if we were to follow the same logic, we will have to show that the new sources we found 
also satisfy the aforementioned conditions. 

 The analysis in Section (\ref{susybranesdef}) showed that D4 branes are no longer supersymmetric embeddings in the deformed geometry. However the new $\tilde{\text{D}}6$ brane sources 
 are provided we turn on a world-volume flux and in what follows we will show that the corrected Bianchi identity associated to them is properly solved. 
 
 For a $\tilde{\text{D}}6$ brane source with world-volume $\text{Mink}_2\times\mathbb{R}^{3}\times T^2$ and localised at $\rho=\rho_0$, the modified Bianchi identity,
  after including world-volume flux, reads
 \begin{equation}\label{bidef}
 d\hat{\tilde{f}}\Big\vert_{(z_7,z_{10})}=(2\pi)^3 N_{D6}\delta(\rho-\rho_0) e^{ \tilde{ f}_2} d\rho\wedge dz_7\wedge dz_{10}.
\end{equation}
Using equations (\ref{bideformed}) and (\ref{deformedf}), integrating across $\rho_0$, allowing $h'_4$ to change an amount $\Delta h'_4$ whilst keeping $h_4, h_8, h'_8, h''_8$ continuous, we obtain
\begin{equation}\label{bianchid6}
\gamma \Delta h'_4=(2\pi)^3 N_{D6}=(2\pi)^3 \gamma N_{D4}, \quad  \tilde{f_2}=\frac{1}{\gamma} dz_9\wedge dz_8, 
\end{equation}
giving rise to a D6-D4 bound state, in agreement with that we found in Section (\ref{susybranesdef}). 

A similar analysis can be carried for the D8. Since the deformation did not affect this source brane,  the analysis follows from the one of the undeformed case. Namely 
the modified Bianchi identity is exactly solved for $\tilde{f}_2=0$.

\section{Killing Spinors on \texorpdfstring{$\mathbb{CP}^3$}{CP3}}\label{app:redCP3}

The solutions presented in Section~\ref{susy06} were classified using bispinor techniques, as developed in~\cite{Macpherson:2023cbl}. This classification is enabled by the fact that the seven-sphere $S^7$ admits a parametrization as an $\mathrm{Sp}(2)$ bundle over $\mathrm{Sp}(1)$.  
Moreover, $S^7$ can be expressed as a fibration over $\mathbb{CP}^3$, with $\mathbb{CP}^3$ itself realized as an $S^2$ fibration over $S^4$. This parametrization was employed in~\cite{Macpherson:2023cbl} to construct AdS$_3$ vacua of type II supergravity that preserve an $\mathrm{SO}(5)$ and $\mathrm{SO}(6)$ isometry. However, for the purposes of the present work—particularly for implementing TsT deformations—we require a more convenient representation of $\mathbb{CP}^3$. Specifically, we adopt a parametrization in which $\mathbb{CP}^3$ is described as a foliation of a $T^{1,1}$ space over an interval. Inspired by~\cite{Legramandi:2020txf}, we use this foliation to explicitly construct the seed bispinors, following the procedure of~\cite{Macpherson:2023cbl}, which will allow us to study the supersymmetry and geometric structure of the deformed backgrounds.

We begin by writing $S^7$ as a foliation of $S^3 \times S^3$ over an interval:
\begin{equation}
	\label{eq:S7met}
	ds^2(S^7) = d\xi^2 + \cos^2\xi \; ds^2(S_1^3) + \sin^2\xi \; ds^2(S_2^3).
\end{equation}
 The geometry admits two sets of Killing spinors, $\xi_\pm$, satisfying
\begin{equation}
	\label{eq:7-sphereKSE}
	\nabla_a \xi_\pm = \pm \frac{i}{2} \gamma_a \xi_\pm.
\end{equation}
We employ the vielbein adapted to this foliation, as given in~\cite{Legramandi:2020txf}:
\begin{align}
	\label{eq:frameS7}
	e^1 &= d\xi, \qquad e^2 = \tfrac{\cos\xi}{2} d\theta_1, \qquad e^3 = \tfrac{\cos\xi}{2} \sin\theta_1\, d\phi_1, \nonumber \\
	e^4 &= \tfrac{\sin\xi}{2} d\theta_2, \qquad e^5 = \tfrac{\sin\xi}{2} \sin\theta_2\, d\phi_2, \nonumber \\
	e^6 &= -\tfrac{\cos\xi\, \sin\xi}{2} \left(d[\psi_1 - \psi_2] + \eta_1 - \eta_2\right), \nonumber \\
	e^7 &= \tfrac{1}{2} \left(\cos^2\xi\, (d\psi_1 + \eta_1) + \sin^2\xi\, (d\psi_2 + \eta_2)\right),
\end{align}
with $\eta_1$ and $\eta_2$ the one-forms on the respective $S^2$ subspaces.
The flat-space gamma matrices are defined as:
$\gamma_1 = \sigma_1 \otimes \mathbb{I}_2 \otimes \mathbb{I}_2$, $\gamma_{2,3,4} = \sigma_2 \otimes \sigma_{1,2,3} \otimes \mathbb{I}_2$, $\gamma_{5,6,7} = \sigma_3 \otimes \mathbb{I}_2 \otimes \sigma_{1,2,3}$, where $\sigma_i$ are the usual Pauli matrices.

The general solution to~\eqref{eq:7-sphereKSE} takes the form
\begin{align}
	\label{eq:squashedS7spinor}
	\xi_\pm &= \mathcal{M}_\pm \, \xi^0_\pm, \\
	\mathcal{M}_\pm &= e^{\frac{\xi}{2}(\pm i \gamma_1 - \gamma_{67})}
	e^{\frac{\psi_1}{4}(\pm i \gamma_7 - \gamma_{23})}
	e^{\frac{\theta_1}{4}(\pm i \gamma_2 + \gamma_{37})}
	e^{\frac{\phi_1}{4}(\pm i \gamma_7 + \gamma_{23})} \nonumber \\
	&\quad \times e^{\frac{\psi_2}{4}(\gamma_{16} - \gamma_{45})}
	e^{\frac{\theta_2}{4}(\gamma_{14} + \gamma_{56})}
	e^{\frac{\phi_2}{4}(\gamma_{16} + \gamma_{45})}, \nonumber
\end{align}
where $\xi^0_\pm$ are constant spinors. One can then construct eight independent spinors as
\begin{equation}
	\xi^A_+ = \mathcal{M}_+ \eta^A, \qquad A = 1, \dots, 8,
\end{equation}
with $\eta^A$ a basis for the $\mathbf{8}$ of $\mathfrak{so}(8)$.

Crucially, the $S^7$ metric in~\eqref{eq:S7met} can also be expressed as a Hopf fibration over $\mathbb{CP}^3$:
\begin{equation}
	ds^2(S^7) = ds^2(\mathbb{CP}^3) + \frac{1}{16} \left(d\tilde{\psi} + \cos^2\xi(d\psi + 2\eta_1) + \sin^2\xi(d\psi - 2\eta_2)\right)^2,
\end{equation}
where we introduced new angular coordinates
\begin{equation}
	\tilde{\psi} = \tfrac{1}{2}(\psi_1 + \psi_2), \qquad \psi = \tfrac{1}{2}(\psi_1 - \psi_2).
\end{equation}
Reducing along the $\tilde{\psi}$ direction yields the $\mathbb{CP}^3$ metric, now realized as a foliation of $T^{1,1}$ over $\xi$. This parametrization is particularly suited for TsT deformations, as it highlights two natural commuting isometries the two $U(1)$'s in S$^2_i$.

The spinors that are uncharged under translations along $\tilde{\psi}$ descend to Killing spinors on $\mathbb{CP}^3$. These span an SO(6) sextuplet and are given explicitly in equation~\eqref{eq:the66}, analogous to the construction in~\cite{Macpherson:2023cbl}. 
These spinors form the basis for constructing the ${\cal N} = (6,0)$ AdS$_3$ solutions we investigate throughout this work.

\vspace{15 pt}
The SU(2)-invariant forms presented in Section~\ref{sec:GSN=2} may be of interest to the reader, as they naturally emerge from the deformation of the original SU(3)-structure bilinears. They are given by,
\begin{align}\label{eq:jUw}
	j=&\frac{1}{p\sqrt{M}}\left(c_{\theta_2}c_\xi s_{\theta_1}^2(e^1\wedge \hat{e}^5- s_{2\xi}c_\psi \hat{e}^3\wedge \hat{e}^6)+(p_1e^4\wedge \hat{e}^5-p_2c_\xi s_\xi e^2\wedge \hat{e}^5)\right)\nonumber\\
	&-\frac{1}{p}\left(s_{\theta_1}(c_{2\xi}e^2\wedge \hat{e}^3+s_{2\xi}s_{\psi}\hat{e}^3\wedge e^4)+c_{\theta_1}s_\xi(e^1\wedge \hat{e}^3+s_{2\xi}c_\psi \hat{e}^5\wedge \hat{e}^6)\right)\nonumber\\
	&-c_\alpha \hat{e}^3\wedge \hat{e}^5+\frac{c_\alpha}{{s_\alpha^2}\sqrt{M}}(c_\xi \;p_3(e^1\wedge e^2- c_{\alpha}e^4\wedge \hat{e}^6)-s_\xi\; p_4\sqrt{M}(e^1\wedge e^4+c_\alpha e^2\wedge \hat{e}^6)\nonumber\\
	&-s_{\theta_1}s_{\theta_2}\sqrt{M}(1-s_{2\xi}^2c_{\psi}^2)e^2\wedge e^4-s_{2\xi}c_\psi(c_{\theta_2}^2s_{\theta_1}^2c_{\xi}^2+c_{\theta_1}^2s_{\theta_2}^2s_{\xi}^2)e^1\wedge \hat{e}^6),\nonumber\\
	U=&\frac{1}{M s_\alpha^2}\left(s_{\theta_1}s_{\theta_2}s_{\psi}e^1-s_\xi(2c_{\theta_2}s_{\theta_1}c_\xi^2+c_{\theta_1}s_{\theta_2}s_\psi c_{2\xi})e^2+c_\xi(2c_{\theta_1}s_{\theta_2}s_\xi^2-c_{\theta_2}s_{\theta_1}s_\psi c_{2\xi})e^4-c_\psi c_{2\xi}\sqrt{M}\hat{e}^6)\right.\nonumber\\
	&\left.+i(c_\psi s_{\theta_1}s_{\theta_2}c_{2\xi}e^1-c_{\theta_1}c_\psi s_{\theta_2}s_\xi e^2-c_{\theta_2}c_\xi c_\psi s_{\theta_1}e^4+s_\psi\sqrt{M}\hat{e}^6)\right),\nonumber\\
	\omega=&\frac{1}{2ps_\alpha\sqrt{M}}\left[2c_{\theta_2}c_\xi s_{\theta_1}^2(e^1\wedge \hat{e}^3+s_{2\xi}c_\psi \hat{e}^5\wedge \hat{e}^6)+s_{2\xi}(2s_{\theta_2} s_\psi(1-c_{\theta_1}^2c_\xi^2)-c_{\theta_1}c_{\theta_2}s_{\theta_1}c_{2\xi})e^2\wedge \hat{e}^3\right.\nonumber\\
	&-(2s_{\theta_2}c_{2\xi}(1-c_{\theta_1}^2c_\xi^2)+c_{\theta_1}c_{\theta_2}s_{\theta_1}s_{2\xi}^2s_\psi)\hat{e}^3\wedge e^4+2\sqrt{M}(c_{\theta_1}s_{\xi}(e^1\wedge \hat{e}^5-s_{2\xi}c_{\psi}\hat{e}^3\wedge \hat{e}^6)\nonumber\\
	&\left.+s_{\theta_1}(c_{2\xi}e^2\wedge \hat{e}^5-s_{2\xi}s_{\psi}e^4\wedge \hat{e}^5))\right]+i\left[-s_\alpha \hat{e}^3\wedge \hat{e}^5+\frac{\cot\alpha}{p}(c_{\theta_1}s_\xi e^1\wedge \hat{e}^3+s_{\theta_1}c_{2\xi}e^2\wedge \hat{e}^3\right.\nonumber\\
	&+s_{\theta_1}s_{2\xi}s_\psi \hat{e}^3\wedge e^4)+\frac{\cot\alpha}{2p\sqrt{M}}(s_{2\xi}(c_{\theta_1}c_{\theta_2}s_{\theta_1}c_{2\xi}-2s_{\theta_2}s_\psi(1-c_{\theta_1}^2c_\xi^2))e^2\wedge \hat{e}^5\nonumber\\
	&+2s_{2\xi}c_\psi(c_{\theta_2}c_\xi s_{\theta_1}^2\hat{e}^3\wedge \hat{e}^6+c_{\theta_1}s_\xi\sqrt{M} \hat{e}^5\wedge \hat{e}^6)-2s_{\theta_1}^2c_\xi c_{\theta_2}e^1\wedge \hat{e}^5-(2s_{\theta_2}c_{2\xi}(1-c_{\theta_1}^2c_{\xi}^2)\nonumber\\
	&+s_{2\xi}^2c_{\theta_1}c_{\theta_2}s_{\theta_1}s_\psi)e^4\wedge \hat{e}^5)-\frac{1}{Ms_\alpha}(s_{2\xi}c_\psi(c_{\theta_2}^2c_{\xi}^2s_{\theta_1}^2+s_{\theta_1}^2s_{\theta_2}^2s_{\psi}^2)e^1\wedge \hat{e}^6+s_{\theta_1}s_{\theta_2}\sqrt{M}(1-c_\psi^2s_{2\xi}^2)e^2\wedge e^4)\nonumber\\
	&\left.+\frac{1}{s_\alpha\sqrt{H}}(p_3 c_\xi (e^1\wedge e^2-c_\alpha e^4\wedge \hat{e}^6)-p_4 s_\xi (e^1\wedge e^4+c_\alpha e^2\wedge \hat{e}^6))\right],\\
	p=&\sqrt{s_\xi^2c_{\theta_1}^2+s_{\theta_1}^2},\qquad M=c_{\theta_2}^2c_{\xi}^2s_{\theta_1}^2+s_{\theta_2}^2p^2,\nonumber\\
	p_1=&\frac{s_{2\xi}^2c_{\theta_1}s_{\theta_1}c_{\theta_2}s_{\psi}}{2}+c_{2\xi}s_{\theta_2}p^2,\qquad 
	p_2=c_{\theta_1}s_{\theta_1}c_{\theta_2}c_{2\xi}-2s_{\theta_2}s_{\psi}p^2,\nonumber\\
	p_3=&c_{\theta_2}s_{\theta_1}c_{2\xi}-2c_{\theta_1}s_{\theta_2}s_{\psi}s_{\xi}^2,\qquad p_4=c_{\theta_1}s_{\theta_2}c_{2\xi}+2c_{\theta_2}s_{\theta_1}s_{\psi}c_{\xi}^2.
\end{align}

\newpage
\bibliography{Bibliography}

\providecommand{\href}[2]{#2}\begingroup\raggedright\begin{thebibliography}{10}

\bibitem{Giveon:1998ns}
A.~Giveon, D.~Kutasov, and N.~Seiberg, ``{Comments on string theory on
  AdS(3)},'' \href{http://dx.doi.org/10.4310/ATMP.1998.v2.n4.a3}{{\em Adv.
  Theor. Math. Phys.} {\bfseries 2} (1998) 733--782},
  \href{http://arxiv.org/abs/hep-th/9806194}{{\ttfamily arXiv:hep-th/9806194}}.

\bibitem{deBoer:1998kjm}
J.~de~Boer, ``{Six-dimensional supergravity on S**3 x AdS(3) and 2-D conformal
  field theory},'' \href{http://dx.doi.org/10.1016/S0550-3213(99)00160-1}{{\em
  Nucl. Phys. B} {\bfseries 548} (1999) 139--166},
  \href{http://arxiv.org/abs/hep-th/9806104}{{\ttfamily arXiv:hep-th/9806104}}.

\bibitem{Lunin:2000yv}
O.~Lunin and S.~D. Mathur, ``{Correlation functions for M**N / S(N)
  orbifolds},'' \href{http://dx.doi.org/10.1007/s002200100431}{{\em Commun.
  Math. Phys.} {\bfseries 219} (2001) 399--442},
  \href{http://arxiv.org/abs/hep-th/0006196}{{\ttfamily arXiv:hep-th/0006196}}.

\bibitem{Maldacena:1999bp}
J.~M. Maldacena, G.~W. Moore, and A.~Strominger, ``{Counting BPS black holes in
  toroidal Type II string theory},''
  \href{http://arxiv.org/abs/hep-th/9903163}{{\ttfamily arXiv:hep-th/9903163}}.

\bibitem{Giveon:1999zm}
A.~Giveon, D.~Kutasov, and O.~Pelc, ``{Holography for noncritical
  superstrings},'' \href{http://dx.doi.org/10.1088/1126-6708/1999/10/035}{{\em
  JHEP} {\bfseries 10} (1999) 035},
  \href{http://arxiv.org/abs/hep-th/9907178}{{\ttfamily arXiv:hep-th/9907178}}.

\bibitem{Gaberdiel:2010pz}
M.~R. Gaberdiel and R.~Gopakumar, ``{An AdS$_{3}$ Dual for Minimal Model
  CFTs},'' \href{http://dx.doi.org/10.1103/PhysRevD.83.066007}{{\em Phys. Rev.
  D} {\bfseries 83} (2011) 066007},
  \href{http://arxiv.org/abs/1011.2986}{{\ttfamily arXiv:1011.2986 [hep-th]}}.

\bibitem{Gukov:2004ym}
S.~Gukov, E.~Martinec, G.~W. Moore, and A.~Strominger, ``{The Search for a
  holographic dual to AdS(3) x S**3 x S**3 x S**1},''
  \href{http://dx.doi.org/10.4310/ATMP.2005.v9.n3.a3}{{\em Adv. Theor. Math.
  Phys.} {\bfseries 9} (2005) 435--525},
  \href{http://arxiv.org/abs/hep-th/0403090}{{\ttfamily arXiv:hep-th/0403090}}.

\bibitem{Couzens:2021veb}
C.~Couzens, Y.~Lozano, N.~Petri, and S.~Vandoren, ``{N=(0,4) black string
  chains},'' \href{http://dx.doi.org/10.1103/PhysRevD.105.086015}{{\em Phys.
  Rev. D} {\bfseries 105} no.~8, (2022) 086015},
  \href{http://arxiv.org/abs/2109.10413}{{\ttfamily arXiv:2109.10413
  [hep-th]}}.

\bibitem{Haghighat:2015ega}
B.~Haghighat, S.~Murthy, C.~Vafa, and S.~Vandoren, ``{F-Theory, Spinning Black
  Holes and Multi-string Branches},''
  \href{http://dx.doi.org/10.1007/JHEP01(2016)009}{{\em JHEP} {\bfseries 01}
  (2016) 009}, \href{http://arxiv.org/abs/1509.00455}{{\ttfamily
  arXiv:1509.00455 [hep-th]}}.

\bibitem{Couzens:2019wls}
C.~Couzens, H.~het Lam, K.~Mayer, and S.~Vandoren, ``{Black Holes and (0,4)
  SCFTs from Type IIB on K3},''
  \href{http://dx.doi.org/10.1007/JHEP08(2019)043}{{\em JHEP} {\bfseries 08}
  (2019) 043}, \href{http://arxiv.org/abs/1904.05361}{{\ttfamily
  arXiv:1904.05361 [hep-th]}}.

\bibitem{Strominger:1996sh}
A.~Strominger and C.~Vafa, ``{Microscopic origin of the Bekenstein-Hawking
  entropy},'' \href{http://dx.doi.org/10.1016/0370-2693(96)00345-0}{{\em Phys.
  Lett. B} {\bfseries 379} (1996) 99--104},
  \href{http://arxiv.org/abs/hep-th/9601029}{{\ttfamily arXiv:hep-th/9601029}}.

\bibitem{Benini:2013cda}
F.~Benini and N.~Bobev, ``{Two-dimensional SCFTs from wrapped branes and
  c-extremization},'' \href{http://dx.doi.org/10.1007/JHEP06(2013)005}{{\em
  JHEP} {\bfseries 06} (2013) 005},
  \href{http://arxiv.org/abs/1302.4451}{{\ttfamily arXiv:1302.4451 [hep-th]}}.

\bibitem{Bah:2019jts}
I.~Bah, F.~Bonetti, R.~Minasian, and E.~Nardoni, ``{Anomaly Inflow for
  M5-branes on Punctured Riemann Surfaces},''
  \href{http://dx.doi.org/10.1007/JHEP06(2019)123}{{\em JHEP} {\bfseries 06}
  (2019) 123}, \href{http://arxiv.org/abs/1904.07250}{{\ttfamily
  arXiv:1904.07250 [hep-th]}}.

\bibitem{Couzens:2018wnk}
C.~Couzens, J.~P. Gauntlett, D.~Martelli, and J.~Sparks, ``{A geometric dual of
  $c$-extremization},'' \href{http://dx.doi.org/10.1007/JHEP01(2019)212}{{\em
  JHEP} {\bfseries 01} (2019) 212},
  \href{http://arxiv.org/abs/1810.11026}{{\ttfamily arXiv:1810.11026
  [hep-th]}}.

\bibitem{Couzens:2022agr}
C.~Couzens, N.~T. Macpherson, and A.~Passias, ``{On Type IIA AdS$_{3}$
  solutions and massive GK geometries},''
  \href{http://dx.doi.org/10.1007/JHEP08(2022)095}{{\em JHEP} {\bfseries 08}
  (2022) 095}, \href{http://arxiv.org/abs/2203.09532}{{\ttfamily
  arXiv:2203.09532 [hep-th]}}.

\bibitem{DHoker:2008rje}
E.~D'Hoker, J.~Estes, M.~Gutperle, and D.~Krym, ``{Exact Half-BPS Flux
  Solutions in M-theory II: Global solutions asymptotic to AdS(7) x S**4},''
  \href{http://dx.doi.org/10.1088/1126-6708/2008/12/044}{{\em JHEP} {\bfseries
  12} (2008) 044}, \href{http://arxiv.org/abs/0810.4647}{{\ttfamily
  arXiv:0810.4647 [hep-th]}}.

\bibitem{DHoker:2009lky}
E.~D'Hoker, J.~Estes, M.~Gutperle, and D.~Krym, ``{Janus solutions in
  M-theory},'' \href{http://dx.doi.org/10.1088/1126-6708/2009/06/018}{{\em
  JHEP} {\bfseries 06} (2009) 018},
  \href{http://arxiv.org/abs/0904.3313}{{\ttfamily arXiv:0904.3313 [hep-th]}}.

\bibitem{Faedo:2020nol}
F.~Faedo, Y.~Lozano, and N.~Petri, ``{Searching for surface defect CFTs within
  AdS$_3$},'' \href{http://dx.doi.org/10.1007/JHEP11(2020)052}{{\em JHEP}
  {\bfseries 11} (2020) 052}, \href{http://arxiv.org/abs/2007.16167}{{\ttfamily
  arXiv:2007.16167 [hep-th]}}.

\bibitem{Lozano:2022ouq}
Y.~Lozano, N.~T. Macpherson, N.~Petri, and C.~Risco, ``{New AdS$_{3}$/CFT$_{2}$
  pairs in massive IIA with (0, 4) and (4, 4) supersymmetries},''
  \href{http://dx.doi.org/10.1007/JHEP09(2022)130}{{\em JHEP} {\bfseries 09}
  (2022) 130}, \href{http://arxiv.org/abs/2206.13541}{{\ttfamily
  arXiv:2206.13541 [hep-th]}}.

\bibitem{Anabalon:2022fti}
A.~Anabal\'on, M.~Chamorro-Burgos, and A.~Guarino, ``{Janus and Hades in
  M-theory},'' \href{http://dx.doi.org/10.1007/JHEP11(2022)150}{{\em JHEP}
  {\bfseries 11} (2022) 150}, \href{http://arxiv.org/abs/2207.09287}{{\ttfamily
  arXiv:2207.09287 [hep-th]}}.

\bibitem{Macpherson:2023cbl}
N.~T. Macpherson and A.~Ramirez, ``{AdS$_{3}$ vacua realising $ \mathfrak{osp}
  $(n|2) superconformal symmetry},''
  \href{http://dx.doi.org/10.1007/JHEP08(2023)024}{{\em JHEP} {\bfseries 08}
  (2023) 024}, \href{http://arxiv.org/abs/2304.12207}{{\ttfamily
  arXiv:2304.12207 [hep-th]}}.

\bibitem{Tong:2014yna}
D.~Tong, ``{The holographic dual of $AdS_{3} \times S^{3} \times S^{3} \times
  S^{1}$},'' \href{http://dx.doi.org/10.1007/JHEP04(2014)193}{{\em JHEP}
  {\bfseries 04} (2014) 193}, \href{http://arxiv.org/abs/1402.5135}{{\ttfamily
  arXiv:1402.5135 [hep-th]}}.

\bibitem{Lozano:2015cra}
Y.~Lozano, N.~T. Macpherson, and J.~Montero, ``{A $ \mathcal{N}=2 $
  supersymmetric AdS$_{4}$ solution in M-theory with purely magnetic flux},''
  \href{http://dx.doi.org/10.1007/JHEP10(2015)004}{{\em JHEP} {\bfseries 10}
  (2015) 004}, \href{http://arxiv.org/abs/1507.02660}{{\ttfamily
  arXiv:1507.02660 [hep-th]}}.

\bibitem{Lozano:2015bra}
Y.~Lozano, N.~T. Macpherson, J.~Montero, and E.~O. Colg\'ain, ``{New $AdS_3
  \times S^2$ T-duals with $ \mathcal{N}=\left(0,4\right) $ supersymmetry},''
  \href{http://dx.doi.org/10.1007/JHEP08(2015)121}{{\em JHEP} {\bfseries 08}
  (2015) 121}, \href{http://arxiv.org/abs/1507.02659}{{\ttfamily
  arXiv:1507.02659 [hep-th]}}.

\bibitem{Kelekci:2016uqv}
O.~Kelekci, Y.~Lozano, J.~Montero, E.~O. Colg\'ain, and M.~Park, ``{Large
  superconformal near-horizons from M-theory},''
  \href{http://dx.doi.org/10.1103/PhysRevD.93.086010}{{\em Phys. Rev. D}
  {\bfseries 93} no.~8, (2016) 086010},
  \href{http://arxiv.org/abs/1602.02802}{{\ttfamily arXiv:1602.02802
  [hep-th]}}.

\bibitem{Couzens:2017way}
C.~Couzens, C.~Lawrie, D.~Martelli, S.~Schafer-Nameki, and J.-M. Wong,
  ``{F-theory and AdS$_{3}$/CFT$_{2}$},''
  \href{http://dx.doi.org/10.1007/JHEP08(2017)043}{{\em JHEP} {\bfseries 08}
  (2017) 043}, \href{http://arxiv.org/abs/1705.04679}{{\ttfamily
  arXiv:1705.04679 [hep-th]}}.

\bibitem{Eberhardt:2017pty}
L.~Eberhardt, M.~R. Gaberdiel, and W.~Li, ``{A holographic dual for string
  theory on
  AdS$_{3}$\texttimes{}S$^{3}$\texttimes{}S$^{3}$\texttimes{}S$^{1}$},''
  \href{http://dx.doi.org/10.1007/JHEP08(2017)111}{{\em JHEP} {\bfseries 08}
  (2017) 111}, \href{http://arxiv.org/abs/1707.02705}{{\ttfamily
  arXiv:1707.02705 [hep-th]}}.

\bibitem{Dibitetto:2017tve}
G.~Dibitetto and N.~Petri, ``{BPS objects in D = 7 supergravity and their
  M-theory origin},'' \href{http://dx.doi.org/10.1007/JHEP12(2017)041}{{\em
  JHEP} {\bfseries 12} (2017) 041},
  \href{http://arxiv.org/abs/1707.06152}{{\ttfamily arXiv:1707.06152
  [hep-th]}}.

\bibitem{Dibitetto:2017klx}
G.~Dibitetto and N.~Petri, ``{6d surface defects from massive type IIA},''
  \href{http://dx.doi.org/10.1007/JHEP01(2018)039}{{\em JHEP} {\bfseries 01}
  (2018) 039}, \href{http://arxiv.org/abs/1707.06154}{{\ttfamily
  arXiv:1707.06154 [hep-th]}}.

\bibitem{Datta:2017ert}
S.~Datta, L.~Eberhardt, and M.~R. Gaberdiel, ``{Stringy $\mathcal{N}=(2,2)$
  holography for AdS${_3}$},''
  \href{http://dx.doi.org/10.1007/JHEP01(2018)146}{{\em JHEP} {\bfseries 01}
  (2018) 146}, \href{http://arxiv.org/abs/1709.06393}{{\ttfamily
  arXiv:1709.06393 [hep-th]}}.

\bibitem{Couzens:2017nnr}
C.~Couzens, D.~Martelli, and S.~Schafer-Nameki, ``{F-theory and
  AdS$_{3}$/CFT$_{2}$ (2, 0)},''
  \href{http://dx.doi.org/10.1007/JHEP06(2018)008}{{\em JHEP} {\bfseries 06}
  (2018) 008}, \href{http://arxiv.org/abs/1712.07631}{{\ttfamily
  arXiv:1712.07631 [hep-th]}}.

\bibitem{Gaberdiel:2018rqv}
M.~R. Gaberdiel and R.~Gopakumar, ``{Tensionless string spectra on
  AdS$_{3}$},'' \href{http://dx.doi.org/10.1007/JHEP05(2018)085}{{\em JHEP}
  {\bfseries 05} (2018) 085}, \href{http://arxiv.org/abs/1803.04423}{{\ttfamily
  arXiv:1803.04423 [hep-th]}}.

\bibitem{Eberhardt:2018sce}
L.~Eberhardt and I.~G. Zadeh, ``{$\mathcal{N}=(3,3)$ holography on ${\rm AdS}_3
  \times ({\rm S}^3 \times {\rm S}^3 \times {\rm S}^1)/\mathbb Z_2$},''
  \href{http://dx.doi.org/10.1007/JHEP07(2018)143}{{\em JHEP} {\bfseries 07}
  (2018) 143}, \href{http://arxiv.org/abs/1805.09832}{{\ttfamily
  arXiv:1805.09832 [hep-th]}}.

\bibitem{Dibitetto:2018ftj}
G.~Dibitetto, G.~Lo~Monaco, A.~Passias, N.~Petri, and A.~Tomasiello, ``{AdS$_3$
  Solutions with Exceptional Supersymmetry},''
  \href{http://dx.doi.org/10.1002/prop.201800060}{{\em Fortsch. Phys.}
  {\bfseries 66} no.~10, (2018) 1800060},
  \href{http://arxiv.org/abs/1807.06602}{{\ttfamily arXiv:1807.06602
  [hep-th]}}.

\bibitem{Dibitetto:2018iar}
G.~Dibitetto and N.~Petri, ``{Surface defects in the D4 $-$ D8 brane system},''
  \href{http://dx.doi.org/10.1007/JHEP01(2019)193}{{\em JHEP} {\bfseries 01}
  (2019) 193}, \href{http://arxiv.org/abs/1807.07768}{{\ttfamily
  arXiv:1807.07768 [hep-th]}}.

\bibitem{Eberhardt:2018ouy}
L.~Eberhardt, M.~R. Gaberdiel, and R.~Gopakumar, ``{The Worldsheet Dual of the
  Symmetric Product CFT},''
  \href{http://dx.doi.org/10.1007/JHEP04(2019)103}{{\em JHEP} {\bfseries 04}
  (2019) 103}, \href{http://arxiv.org/abs/1812.01007}{{\ttfamily
  arXiv:1812.01007 [hep-th]}}.

\bibitem{Macpherson:2018mif}
N.~T. Macpherson, ``{Type II solutions on AdS$_{3} \times$ S$^{3} \times$
  S$^{3}$ with large superconformal symmetry},''
  \href{http://dx.doi.org/10.1007/JHEP05(2019)089}{{\em JHEP} {\bfseries 05}
  (2019) 089}, \href{http://arxiv.org/abs/1812.10172}{{\ttfamily
  arXiv:1812.10172 [hep-th]}}.

\bibitem{Lozano:2019emq}
Y.~Lozano, N.~T. Macpherson, C.~Nunez, and A.~Ramirez, ``{AdS$_3$ solutions in
  Massive IIA with small $\mathcal{N}=(4,0)$ supersymmetry},''
  \href{http://dx.doi.org/10.1007/JHEP01(2020)129}{{\em JHEP} {\bfseries 01}
  (2020) 129}, \href{http://arxiv.org/abs/1908.09851}{{\ttfamily
  arXiv:1908.09851 [hep-th]}}.

\bibitem{Lozano:2019jza}
Y.~Lozano, N.~T. Macpherson, C.~Nunez, and A.~Ramirez, ``{1/4 BPS solutions and
  the AdS$_3$/CFT$_2$ correspondence},''
  \href{http://dx.doi.org/10.1103/PhysRevD.101.026014}{{\em Phys. Rev. D}
  {\bfseries 101} no.~2, (2020) 026014},
  \href{http://arxiv.org/abs/1909.09636}{{\ttfamily arXiv:1909.09636
  [hep-th]}}.

\bibitem{Lozano:2019zvg}
Y.~Lozano, N.~T. Macpherson, C.~Nunez, and A.~Ramirez, ``{Two dimensional
  ${\cal N}=(0,4)$ quivers dual to AdS$_3$ solutions in massive IIA},''
  \href{http://dx.doi.org/10.1007/JHEP01(2020)140}{{\em JHEP} {\bfseries 01}
  (2020) 140}, \href{http://arxiv.org/abs/1909.10510}{{\ttfamily
  arXiv:1909.10510 [hep-th]}}.

\bibitem{Lozano:2019ywa}
Y.~Lozano, N.~T. Macpherson, C.~Nunez, and A.~Ramirez, ``{AdS$_3$ solutions in
  massive IIA, defect CFTs and T-duality},''
  \href{http://dx.doi.org/10.1007/JHEP12(2019)013}{{\em JHEP} {\bfseries 12}
  (2019) 013}, \href{http://arxiv.org/abs/1909.11669}{{\ttfamily
  arXiv:1909.11669 [hep-th]}}.

\bibitem{Passias:2019rga}
A.~Passias and D.~Prins, ``{On AdS$_3$ solutions of Type IIB},''
  \href{http://dx.doi.org/10.1007/JHEP05(2020)048}{{\em JHEP} {\bfseries 05}
  (2020) 048}, \href{http://arxiv.org/abs/1910.06326}{{\ttfamily
  arXiv:1910.06326 [hep-th]}}.

\bibitem{Eberhardt:2019ywk}
L.~Eberhardt, M.~R. Gaberdiel, and R.~Gopakumar, ``{Deriving the
  AdS$_{3}$/CFT$_{2}$ correspondence},''
  \href{http://dx.doi.org/10.1007/JHEP02(2020)136}{{\em JHEP} {\bfseries 02}
  (2020) 136}, \href{http://arxiv.org/abs/1911.00378}{{\ttfamily
  arXiv:1911.00378 [hep-th]}}.

\bibitem{Couzens:2019iog}
C.~Couzens, ``{$ \mathcal{N} $ = (0, 2) AdS$_{3}$ solutions of type IIB and
  F-theory with generic fluxes},''
  \href{http://dx.doi.org/10.1007/JHEP04(2021)038}{{\em JHEP} {\bfseries 04}
  (2021) 038}, \href{http://arxiv.org/abs/1911.04439}{{\ttfamily
  arXiv:1911.04439 [hep-th]}}.

\bibitem{Couzens:2019mkh}
C.~Couzens, H.~het Lam, and K.~Mayer, ``{Twisted $ \mathcal{N} $ = 1 SCFTs and
  their AdS$_{3}$ duals},''
  \href{http://dx.doi.org/10.1007/JHEP03(2020)032}{{\em JHEP} {\bfseries 03}
  (2020) 032}, \href{http://arxiv.org/abs/1912.07605}{{\ttfamily
  arXiv:1912.07605 [hep-th]}}.

\bibitem{Legramandi:2019xqd}
A.~Legramandi and N.~T. Macpherson, ``{AdS$_3$ solutions with from
  $\mathcal{N}=(3,0)$ from S$^3\times$S$^3$ fibrations},''
  \href{http://dx.doi.org/10.1002/prop.202000014}{{\em Fortsch. Phys.}
  {\bfseries 68} no.~3-4, (2020) 2000014},
  \href{http://arxiv.org/abs/1912.10509}{{\ttfamily arXiv:1912.10509
  [hep-th]}}.

\bibitem{Lozano:2020bxo}
Y.~Lozano, C.~Nunez, A.~Ramirez, and S.~Speziali, ``{$M$-strings and AdS$_3$
  solutions to M-theory with small $\mathcal{N}=(0,4)$ supersymmetry},''
  \href{http://dx.doi.org/10.1007/JHEP08(2020)118}{{\em JHEP} {\bfseries 08}
  (2020) 118}, \href{http://arxiv.org/abs/2005.06561}{{\ttfamily
  arXiv:2005.06561 [hep-th]}}.

\bibitem{Dibitetto:2020bsh}
G.~Dibitetto and N.~Petri, ``{AdS$_{3}$ from M-branes at conical
  singularities},'' \href{http://dx.doi.org/10.1007/JHEP01(2021)129}{{\em JHEP}
  {\bfseries 01} (2021) 129}, \href{http://arxiv.org/abs/2010.12323}{{\ttfamily
  arXiv:2010.12323 [hep-th]}}.

\bibitem{Passias:2020ubv}
A.~Passias and D.~Prins, ``{On supersymmetric AdS$_{3}$ solutions of Type
  II},'' \href{http://dx.doi.org/10.1007/JHEP08(2021)168}{{\em JHEP} {\bfseries
  08} (2021) 168}, \href{http://arxiv.org/abs/2011.00008}{{\ttfamily
  arXiv:2011.00008 [hep-th]}}.

\bibitem{Faedo:2020lyw}
F.~Faedo, Y.~Lozano, and N.~Petri, ``{New $\mathcal{N}=(0,4)$ AdS$_3$
  near-horizons in Type IIB},''
  \href{http://dx.doi.org/10.1007/JHEP04(2021)028}{{\em JHEP} {\bfseries 04}
  (2021) 028}, \href{http://arxiv.org/abs/2012.07148}{{\ttfamily
  arXiv:2012.07148 [hep-th]}}.

\bibitem{Legramandi:2020txf}
A.~Legramandi, G.~Lo~Monaco, and N.~T. Macpherson, ``{All $\mathcal{N}=(8,0)$
  AdS$_3$ solutions in 10 and 11 dimensions},''
  \href{http://dx.doi.org/10.1007/JHEP05(2021)263}{{\em JHEP} {\bfseries 05}
  (2021) 263}, \href{http://arxiv.org/abs/2012.10507}{{\ttfamily
  arXiv:2012.10507 [hep-th]}}.

\bibitem{Couzens:2021tnv}
C.~Couzens, N.~T. Macpherson, and A.~Passias, ``{$ \mathcal{N} $ = (2, 2)
  AdS$_{3}$ from D3-branes wrapped on Riemann surfaces},''
  \href{http://dx.doi.org/10.1007/JHEP02(2022)189}{{\em JHEP} {\bfseries 02}
  (2022) 189}, \href{http://arxiv.org/abs/2107.13562}{{\ttfamily
  arXiv:2107.13562 [hep-th]}}.

\bibitem{Macpherson:2021lbr}
N.~T. Macpherson and A.~Tomasiello, ``{$ \mathcal{N} $ = (1, 1) supersymmetric
  AdS$_{3}$ in 10 dimensions},''
  \href{http://dx.doi.org/10.1007/JHEP03(2022)112}{{\em JHEP} {\bfseries 03}
  (2022) 112}, \href{http://arxiv.org/abs/2110.01627}{{\ttfamily
  arXiv:2110.01627 [hep-th]}}.

\bibitem{Macpherson:2022sbs}
N.~T. Macpherson and A.~Ramirez, ``{AdS$_{3}$\texttimes{}S$^{2}$ in IIB with
  small $ \mathcal{N} $ = (4, 0) supersymmetry},''
  \href{http://dx.doi.org/10.1007/JHEP04(2022)143}{{\em JHEP} {\bfseries 04}
  (2022) 143}, \href{http://arxiv.org/abs/2202.00352}{{\ttfamily
  arXiv:2202.00352 [hep-th]}}.

\bibitem{Conti:2024rqy}
A.~Conti and N.~T. Macpherson, ``{${\cal N}=(4,4)$ supersymmetric AdS$_3$
  solutions in $d=11$},'' \href{http://arxiv.org/abs/2408.17303}{{\ttfamily
  arXiv:2408.17303 [hep-th]}}.

\bibitem{Conti:2024rwd}
A.~Conti, G.~Dibitetto, Y.~Lozano, N.~Petri, and A.~Ram\'\i{}rez, ``{Half-BPS
  Janus solutions in AdS$_{7}$},''
  \href{http://dx.doi.org/10.1007/JHEP12(2024)198}{{\em JHEP} {\bfseries 12}
  (2024) 198}, \href{http://arxiv.org/abs/2407.21619}{{\ttfamily
  arXiv:2407.21619 [hep-th]}}.

\bibitem{Lozano:2024idt}
Y.~Lozano, N.~T. Macpherson, N.~Petri, and A.~Ram\'\i{}rez, ``{Holographic $
  \frac{1}{2} $-BPS surface defects in ABJM},''
  \href{http://dx.doi.org/10.1007/JHEP08(2024)044}{{\em JHEP} {\bfseries 08}
  (2024) 044}, \href{http://arxiv.org/abs/2404.17469}{{\ttfamily
  arXiv:2404.17469 [hep-th]}}.

\bibitem{Lunin:2005jy}
O.~Lunin and J.~M. Maldacena, ``{Deforming field theories with U(1) x U(1)
  global symmetry and their gravity duals},''
  \href{http://dx.doi.org/10.1088/1126-6708/2005/05/033}{{\em JHEP} {\bfseries
  05} (2005) 033}, \href{http://arxiv.org/abs/hep-th/0502086}{{\ttfamily
  arXiv:hep-th/0502086}}.

\bibitem{Gauntlett:2002sc}
J.~P. Gauntlett, D.~Martelli, S.~Pakis, and D.~Waldram, ``{G structures and
  wrapped NS5-branes},''
  \href{http://dx.doi.org/10.1007/s00220-004-1066-y}{{\em Commun. Math. Phys.}
  {\bfseries 247} (2004) 421--445},
  \href{http://arxiv.org/abs/hep-th/0205050}{{\ttfamily arXiv:hep-th/0205050}}.

\bibitem{Gauntlett:2003cy}
J.~P. Gauntlett, D.~Martelli, and D.~Waldram, ``{Superstrings with intrinsic
  torsion},'' \href{http://dx.doi.org/10.1103/PhysRevD.69.086002}{{\em Phys.
  Rev. D} {\bfseries 69} (2004) 086002},
  \href{http://arxiv.org/abs/hep-th/0302158}{{\ttfamily arXiv:hep-th/0302158}}.

\bibitem{Gauntlett:2004zh}
J.~P. Gauntlett, D.~Martelli, J.~Sparks, and D.~Waldram, ``{Supersymmetric
  AdS(5) solutions of M theory},''
  \href{http://dx.doi.org/10.1088/0264-9381/21/18/005}{{\em Class. Quant.
  Grav.} {\bfseries 21} (2004) 4335--4366},
  \href{http://arxiv.org/abs/hep-th/0402153}{{\ttfamily arXiv:hep-th/0402153}}.

\bibitem{Grana:2005sn}
M.~Grana, R.~Minasian, M.~Petrini, and A.~Tomasiello, ``{Generalized structures
  of N=1 vacua},'' \href{http://dx.doi.org/10.1088/1126-6708/2005/11/020}{{\em
  JHEP} {\bfseries 11} (2005) 020},
  \href{http://arxiv.org/abs/hep-th/0505212}{{\ttfamily arXiv:hep-th/0505212}}.

\bibitem{Grana:2006kf}
M.~Grana, R.~Minasian, M.~Petrini, and A.~Tomasiello, ``{A Scan for new N=1
  vacua on twisted tori},''
  \href{http://dx.doi.org/10.1088/1126-6708/2007/05/031}{{\em JHEP} {\bfseries
  05} (2007) 031}, \href{http://arxiv.org/abs/hep-th/0609124}{{\ttfamily
  arXiv:hep-th/0609124}}.

\bibitem{Shelton:2005cf}
J.~Shelton, W.~Taylor, and B.~Wecht, ``{Nongeometric flux compactifications},''
  \href{http://dx.doi.org/10.1088/1126-6708/2005/10/085}{{\em JHEP} {\bfseries
  10} (2005) 085}, \href{http://arxiv.org/abs/hep-th/0508133}{{\ttfamily
  arXiv:hep-th/0508133}}.

\bibitem{Andriot:2011uh}
D.~Andriot, M.~Larfors, D.~Lust, and P.~Patalong, ``{A ten-dimensional action
  for non-geometric fluxes},''
  \href{http://dx.doi.org/10.1007/JHEP09(2011)134}{{\em JHEP} {\bfseries 09}
  (2011) 134}, \href{http://arxiv.org/abs/1106.4015}{{\ttfamily arXiv:1106.4015
  [hep-th]}}.

\bibitem{Filippas:2020qku}
K.~Filippas, ``{Holography for 2D $\mathcal{N}=(0,4)$ quantum field theory},''
  \href{http://dx.doi.org/10.1103/PhysRevD.103.086003}{{\em Phys. Rev. D}
  {\bfseries 103} no.~8, (2021) 086003},
  \href{http://arxiv.org/abs/2008.00314}{{\ttfamily arXiv:2008.00314
  [hep-th]}}.

\bibitem{Apruzzi:2015zna}
F.~Apruzzi, M.~Fazzi, A.~Passias, and A.~Tomasiello, ``{Supersymmetric
  AdS$_{5}$ solutions of massive IIA supergravity},''
  \href{http://dx.doi.org/10.1007/JHEP06(2015)195}{{\em JHEP} {\bfseries 06}
  (2015) 195}, \href{http://arxiv.org/abs/1502.06620}{{\ttfamily
  arXiv:1502.06620 [hep-th]}}.

\bibitem{Aharony:2008ug}
O.~Aharony, O.~Bergman, D.~L. Jafferis, and J.~Maldacena, ``{N=6 superconformal
  Chern-Simons-matter theories, M2-branes and their gravity duals},''
  \href{http://dx.doi.org/10.1088/1126-6708/2008/10/091}{{\em JHEP} {\bfseries
  10} (2008) 091}, \href{http://arxiv.org/abs/0806.1218}{{\ttfamily
  arXiv:0806.1218 [hep-th]}}.

\bibitem{Conti:2025djz}
A.~Conti, Y.~Lozano, and N.~T. Macpherson, ``{$\mathcal{N}=6$ supersymmetric
  AdS$_2 \times \mathbb{CP}^3\times \Sigma_2 $},''
  \href{http://arxiv.org/abs/2503.23585}{{\ttfamily arXiv:2503.23585
  [hep-th]}}.

\bibitem{Zacarias:2021pfz}
S.~Zacarias, ``{Marginal deformations of a class of AdS$_{3} \mathcal{N} $ =
  (0, 4) holographic backgrounds},''
  \href{http://dx.doi.org/10.1007/JHEP06(2021)017}{{\em JHEP} {\bfseries 06}
  (2021) 017}, \href{http://arxiv.org/abs/2102.05681}{{\ttfamily
  arXiv:2102.05681 [hep-th]}}.

\bibitem{Imeroni:2008cr}
E.~Imeroni, ``{On deformed gauge theories and their string/M-theory duals},''
  \href{http://dx.doi.org/10.1088/1126-6708/2008/10/026}{{\em JHEP} {\bfseries
  10} (2008) 026}, \href{http://arxiv.org/abs/0808.1271}{{\ttfamily
  arXiv:0808.1271 [hep-th]}}.

\bibitem{Conti:2023rul}
A.~Conti, ``{AdS3 T duality and evidence for N=5,6 superconformal quantum
  mechanics},'' \href{http://dx.doi.org/10.1103/PhysRevD.108.126007}{{\em Phys.
  Rev. D} {\bfseries 108} no.~12, (2023) 126007},
  \href{http://arxiv.org/abs/2306.09139}{{\ttfamily arXiv:2306.09139
  [hep-th]}}.

\bibitem{Butti:2007aq}
A.~Butti, D.~Forcella, L.~Martucci, R.~Minasian, M.~Petrini, and A.~Zaffaroni,
  ``{On the geometry and the moduli space of beta-deformed quiver gauge
  theories},'' \href{http://dx.doi.org/10.1088/1126-6708/2008/07/053}{{\em
  JHEP} {\bfseries 07} (2008) 053},
  \href{http://arxiv.org/abs/0712.1215}{{\ttfamily arXiv:0712.1215 [hep-th]}}.

\bibitem{Minasian:2006hv}
R.~Minasian, M.~Petrini, and A.~Zaffaroni, ``{Gravity duals to deformed SYM
  theories and Generalized Complex Geometry},''
  \href{http://dx.doi.org/10.1088/1126-6708/2006/12/055}{{\em JHEP} {\bfseries
  12} (2006) 055}, \href{http://arxiv.org/abs/hep-th/0606257}{{\ttfamily
  arXiv:hep-th/0606257}}.

\bibitem{Grana:2008yw}
M.~Grana, R.~Minasian, M.~Petrini, and D.~Waldram, ``{T-duality, Generalized
  Geometry and Non-Geometric Backgrounds},''
  \href{http://dx.doi.org/10.1088/1126-6708/2009/04/075}{{\em JHEP} {\bfseries
  04} (2009) 075}, \href{http://arxiv.org/abs/0807.4527}{{\ttfamily
  arXiv:0807.4527 [hep-th]}}.

\bibitem{Bergshoeff:1994cb}
E.~Bergshoeff, R.~Kallosh, and T.~Ortin, ``{Duality versus supersymmetry and
  compactification},'' \href{http://dx.doi.org/10.1103/PhysRevD.51.3009}{{\em
  Phys. Rev. D} {\bfseries 51} (1995) 3009--3016},
  \href{http://arxiv.org/abs/hep-th/9410230}{{\ttfamily arXiv:hep-th/9410230}}.

\end{thebibliography}\endgroup

\end{document}